\documentclass[prd,showpacs]{revtex4}
\usepackage{caption}
\usepackage{amsmath}
\usepackage{graphicx}
\usepackage{epsfig}
\usepackage{dcolumn}
\usepackage{bm}
\usepackage{slashed}
\usepackage{amsfonts}
\usepackage{amssymb}

\begin{document}
\title{Quantum phase transition from a Antiferromagnetic-Insulator to a Paramagnetic-Metal laying beneath the superconducting dome}
\author{Victor M. Martinez Alvarez$^{*,***}$, Alejandro Cabo-Bizet$^{**}$ and Alejandro Cabo$^{*}$ }

\address{ $^{*}$   Department de F\'{\i}sica Te\'orica,
Instituto de Cibern\'{e}tica, Matem\'{a}tica y F\'{\i}sica, Calle E,
No. 309, Vedado, La Habana, Cuba.}

\address{ $^{**}$ Department de F\'{\i}sica,
Centro de Aplicaciones Tecnol\'ogicas y  Desarrollo Nuclear, Calle 30, No. 502 e/ 5ta y 7ma Avenida,
Playa, La Habana, Cuba.}

\address{ $^{***}$ Department de F\'{\i}sica,
Universidad de Camag\"uey, Circunvalaci\'{o}n Norte, km 5, Camag\"uey, Cuba.}

\begin{abstract}
The effect of hole doping on the Tight-Binding ($TB$) model of the
$Cu$-$O$ planes in the $La_{2}CuO_{4}$ constructed in previous works
is investigated here. Firstly, it is pointed out that the model
employed constitutes a generalization of the Hubbard one for the
same system. Thus, the former predictions of the insulator gap,
antiferromagnetic ($AF$) character  and the existence
of a paramagnetic-pseudogap ($PPG$) state at half-filling, become  natural ones
to be expected from this more general picture. The effect of hole
doping on  the antiferromagnetic-insulator state
($AFI$) and the paramagnetic-pseudogap one at half-filling,
is investigated here at $T=0 \,K$. The results predict the occurrence
of a quantum phase transition ($QPT$) from the antiferromagnetic-
insulator state at low doping to a paramagnetic-metallic state ($PM$)
at higher hole densities. Therefore, a clear description of the
hidden $QPT$ laying beneath the \textquotedblleft
dome\textquotedblright\ in high critical temperature ($HTc$)
superconducting materials is found. At low doping, the systems
prefers the $AFI$ state, and at the critical value of the doping
density $\delta_{c}=0.2 $, the energy of a metallic  state starts
becoming lower. The evolution with  small doping values  of the band spectrum of
the $AFI$ state, shows that the holes tend to become localized at
the middle of the sides of the reduced Brillouin zone ($BZ$). Then, when $\delta$
passes through  the critical value, the holes of the $AFI$ state
move to become situated at the corners of the same reduced $BZ$,
showing in this way a structural change at the phase transition
point. Thus, the  $PM$ state which appears at the transition point
acquires the same behavior with respect to the localization of holes
as the pseudogap state. In the small doping limit a clear difference
between the degree of convergence of the iterative self-consistent
solution is associated to an even or odd number of electrons. It
suggests that the Kramers degeneration in combination with the
spin-spatial entangled nature of the hole states, leads to a new
kind of pair interaction between two holes. The binding energy value is estimated
as a function of the screening.
\end{abstract}
\maketitle

\section{Introduction}

Since  the discovery of the high critical temperature
superconductivity in the called \textquotedblleft
cuprates\textquotedblright\ (made up of copper oxides) \cite{Bednorz},
the search about the nature of their behavior has been and it
continues being, one  of the most defiant  and investigated topics
in Condensed Matter Physics.  Among these compounds  is the
$La_{2}CuO_{4}$, that is a classic  example  of the so-called
strongly correlated electron systems ($SCES$). However, due to the
inherent difficulties of the many body problem for this class of
materials, numerous properties continue without being clearing up,
in spite of the efforts done by the investigators in this subject.
The $La_{2}CuO_{4}$ superconductor is characterized by a drastic  change in its behavior
depending on the density of electrons in the two-dimensional planes
$CuO_{2}$ \cite{Dagotto}, and it  happens in  general in the $HT_c$ superconductivity
materials. All the cuprate superconductors have as a common element,
the existence in their crystal structure of $CuO_{2}$ planes.

The  standard band calculations predict a character of paramagnetic-conductor
to this material \cite{Matheiss}, in drastic contrast with its
experimentally observed, insulating and antiferromagnetic nature.
These two  properties  are  associated with the presence  of strong correlations, and are not
derivable starting from an  independent particle scheme  as the
Hartree-Fock ($HF$) one, when the whole many electron crystal system is considered \cite{Dagotto}. However, in Refs. \cite{Cabo1,Cabo2,Cabo3}, on the base of a single band model solved in the $HF$ approximation, it was possible to  predict the mentioned $SCES$ properties of this material. At first sight, it might seem that to  obtain  the  $AF$ and insulating properties of  $La_2CuO_4$ from a $HF$ scheme, could be  a contradictory outcome,  since those properties  are thought as essentially indicating  the  strong correlation nature of this material.  However, those results  become natural ones, when considering the following circumstances. In first place, the $HF$ procedure was not applied  in its quality of a $First$ $Principle$ method to solve the full Hamiltonian problem  associated to the total electronic structure of  $La_2CuO_4$.  Alternatively, the procedure   was employed for  solving the  simple model of the $CuO_2$  layers built in Refs.  \cite{Cabo1,Cabo2,Cabo3}.  It is well-known that the impossibility of describing strong correlation properties by employing  the $HF$ scheme,  appears when the scheme is   applied to find the solution of the exact many electron  problem defined by all the  electrons and nuclei constituting a crystalline solid. For example, the direct $HF$  calculation of the band structure of $La_2CuO_4$,  predicts an enormous electronic  gap of nearly of $17\,eV$ \cite{Yen}. In addition, the model introduced in Refs. \cite{Cabo1,Cabo2,Cabo3}, only considers the electrons that half-filled the single band crossing the Fermi level appearing in the early electronic band calculation  \cite{Matheiss}. In fact, the proposed  simple Hamiltonian can be viewed as a preliminary stage within the definition of one among the variety of Hubbard's Hamiltonians. One, in which the last step of imposing the nearest neighbor approximations leading to the Hubbard theories had not  been implemented.  Therefore, the proposed model could be considered as an improvement of a Hubbard one, which retains the full Coulomb repulsion operator intact \cite{Fradkin}. This last point can be understood by keeping in mind that its  free Hamiltonian is basically a $TB$ one in which the full Coulomb interaction operator between the electrons had been retained. Thus, after performing the nearest neighbor approximations,  a typical Hubbard scheme  arises.   Henceforth, since the Hubbard model is recognized to convey strong correlation effects,  and moreover, since it is also known  that it’s mean field  solutions also can exhibit such effects \cite{Fradkin}, the appearance  of the insulator gap and the antiferromagnetic structure in Refs. \cite{Cabo1,Cabo2,Cabo3}, becomes a reasonable conclusion. However, a new physical prediction coming from the analysis is the existence of pseudogap states, which are not following in the Hubbard approximation. These mentioned results  emerged as a
consequence of a combination of lattice symmetry breaking with a spin-space entanglement  structure  in the single particle solutions of the $HF$ problem. The treatment gave a clear explanation about the nature of the
antiferromagnetism and the insulating structure of this high
temperature superconductor.

The process through which  the electronic
structure evolves with the hole doping, from the $AFI$ state at low
doping  to the superconductor state, and then to a normal metal
phase at large hole concentrations, has been the subject of a strong interest in the literature. It is known
that the magnitude of the gap in the normal state is of the same order that the superconductor gap \cite{Ino}.
With the objective of studying the hole doping dependence of the
band structure, the $La_{2-x}Sr_{x}CuO_{4}$ ($LSCO$) system  is for
several reasons appropriate among the family of the $HTc$
superconductors. In first place, the $LSCO$ has a simple crystalline
structure of  $CuO_{2}$ layers. Second, the concentration of holes
in these planes can be controlled in a wide range and is uniquely
determined by the concentration $x$ of $Sr$ \cite{Ino}. For this reason,
one can develop samples of the material by varying  doping continuously from
the insulating state  without doping $(x=0)$ up to  the high doping
limit $(x=0.3)$ in the same system.

Numerous experiments made in copper oxide compounds, suggest the
existence of a quantum phase transition at $T=0 \,K$, which lays inside the doping interval of the superconducting dome (this is colloquially referred as   \textquotedblleft laying beneath\textquotedblright\ the dome). It is believed that this $QPT$  is the key to understand the high temperature
superconductivity and also to explain the properties of the normal
state in those materials \cite{Broun,Sachdev}. The present work presents results that  predict the existence of this sort of $QPT$ in the context of the  model for the $La_{2}%
CuO_{4}$ constructed in Refs. \cite{Cabo1,Cabo2,Cabo3} when
the hole doping is incorporated. The $HF$ system of equations  is
solved here by employing the same  method used in Refs.
\cite{Cabo1,Cabo2,Cabo3}. In the present case, again the parameters
of the model are determined of imposing the condition that the $HF$
solution without crystalline symmetry breaking,  reproduces the form
of the single  band crossing the Fermi level in the band calculations
of the material \cite{Matheiss}. Then, we investigated the changes that occur in the band structure when the system is doped with holes. As in Refs. \cite{Cabo1,Cabo2,Cabo3} at exact half-filling, the $HF$ solution predicts the existence of the $AFI$ and $PPG$ states. This previous result  motivated  the idea of investigating the hole doping consequences, after  conceiving the  possibility of predicting the quantum phase transition at $T=0\,K$. The same existence of this transition, is currently one of the most fundamental questions  in high temperature  superconductivity research \cite{Broun,Sachdev}. A transition like this is now  considered that should determine the properties of the normal state in the diverse regions of the phase diagram. The present investigation aim consists in studying the existence of this quantum phase transition beneath the superconducting dome.

The  Hartree-Fock description of the simple model for the
$Cu$-$O$ planes of  $La_{2}CuO_{4}$  including symmetry breaking
effects and spin-spatial  \textquotedblleft
entanglement\textquotedblright\ of the single particle states  at
half-filling,  is here extended to  consider the effects of hole
doping. It is observed how the effect of doping  is able to
predict a variety of the  most interesting properties of this
material a $T=0 \,K$. The evolution of the band spectrum of the
antiferromagnetic-insulator and the paramagnetic-pseudogap states
as functions of the hole doping parameter are determined, for a wide
range of the hole concentration $0\leq x\leq0.3$. Around the
critical doping  $x_{c}=0.2$, the results show that for the  $AFI$ state, the
band spectrum suffers a gradual  change, in which the insulator  gap
diminishes to completely closing. Surprisingly,  the same behavior
occurs for the $PPG$ state for which the pseudogap also collapses.
For higher values than the mentioned  critical doping one, the
$AFI$ and $PPG$ states coalesce in a single $HF$ solution. The
magnetization of this high doping  state vanishes as well as its
gap, thus, it describes a paramagnetic-metallic phase. The results
also reveal a drastic change in the Fermi surface, which goes
from a hole-like Fermi surface centered at ($\pi,\pi$) for $0<x<0.2$
into an electron-like one centered at ($0,0$) for $0.2<x\leq0.3$. These results evidence the existence of a quantum phase transition laying beneath the superconductor dome,
in which an insulator state with antiferromagnetic correlations
transits to a paramagnetic-metallic phase. The work also identifies a
new mechanism of hole pairing which could give rise to the
superconductivity. The effect results from a combination of the
Kramers degeneration with the spin-spatial entanglement of the
single particle $HF$ states. The discussion helps to clarify the
relation between: the pair of states $AFI$ and $PPG$ in the
Physics of the strongly correlated electron systems with the
superconductor state and the quantum critical point, around which
the phase transition occurs \cite{Broun,Sachdev,Tallon,loram}. An evaluation of the hole pairs binding energy as
a function of the dielectric constant is obtained, which predicts binding of the pairs at the experimentally measured values of the dielectric constant of $La_2CuO_4$ \cite{Chen}.
However, the thermodynamic limit for this calculation had not been attained yet in this work.
In addition, the parameters of the model should  be optimized yet to match the measured
parameters of the material as the insulator gap of $2\,eV$ \cite{Ginder}, and the dielectric constant having a value
of the order of $25$ \cite{Chen}.  Here, it is pointed out that this  binding effect could be the acting mechanism  determining the physical  relevance of the doubly charged  bosonic  fields argued  in Ref. \cite{Philips}.

The papers is organized as follows. In section 2 we  review the one band model of the $Cu$-$O$ planes in $La_{2}CuO_{4}$ introduced in Refs. \cite{Cabo1,Cabo2,Cabo3}, the procedure for the
determination of its parameters  and the Hartree-Fock solution for the here considered situation: the study of the hole doping effects. In section 3 the results are presented for the evolution of the band spectrum of the $PPG$ state in the range of doping $0\leq x\leq0.3$. Next, the section 4 considers the same study for the $AFI$ state. Section 5  presents the results for the changes of the Fermi surfaces as the doping is increased. The quantum phase transition properties are discussed in section 6. At the section 7 we exposes the identification of a possible hole pairing mechanisms which could give rise to the superconductivity and estimate the values of the energy for pair binding. Finally, the conclusion are presented in section 8.

\section{The $CuO_{2}$ model including hole doping }

In this section we will review the main ideas and elements defining
the model for the $La_{2}CuO_{4}$ introduced in Refs. \cite{Cabo1,Cabo2,Cabo3}. In the first subsection, the structure and notation of the unrestricted Hartree-Fock scheme employed is
described. Next, the model is presented.
\subsection{Fully unrestricted Hartree-Fock scheme }

The $N$ electron system considered in the $TB$ model discussed in
Refs. \cite{Cabo1,Cabo2,Cabo3} was  described by a fully
unrestricted Slater determinant $f_{n}(x_{1};s_{1},...,x_{N};s_{N})$
state constructed with  single particle orbitals
$\phi_{k_{i}}(x_{i},s_{i})$ with $i=1,...,N$, which arbitrarily
depends on the spin variables at any point of the space. The index
$n$ represents the set of quantum numbers of the many electron
system. In this subsection, as usual,  generalized coordinates will be assumed to
incorporate the spatial in common with the spin ones. For the
electron case under study, the Slater determinant gets the expression
\begin{equation}
\begin{array}
[c]{ccc}%
f(x_{1};s_{1},...,x_{N};s_{N}) & = & \frac{1}{\sqrt{N!}}\sum_{\eta
_{1},...,\eta_{N}}\epsilon^{\eta_{1},...,\eta_{N}}\phi_{\eta_{1}}(x_{1}%
,s_{1})...\phi_{\eta_{N}}(x_{N},s_{N}),\\
&  & \forall_{i}\,\,\eta_{i}=k_{1},...,k_{N},
\end{array}
\label{eq:1}
\end{equation}
where  $\epsilon^{\eta_{1},...,\eta_{N}}$  is the  Levi-Civita
tensor. The single particle $HF$ orbitals satisfy a set of coupled
integro-differential equations of the Pauli kind. This set  is derived from the minimization of the $HF$ energy functional of the system under the conditions of normalization of
$f$ and the normalization of all the single particle orbitals
$\phi_{k_{i}}$. For more details see Refs. \cite{Cabo1,Cabo2,Cabo3}.
In general, the  Hamiltonian of  usual electronic systems has a free
term $\sum_{i}\hat{\mathcal{H}}_{0}(x_{i})$ (a kinetic energy plus
an interaction with an external field one) with an addition
associated to the pair Coulomb interaction between the electrons in
the form
\begin{equation}
\hat{\mathcal{H}}(x_{1},...,x_{N})=\sum_{i}\hat{\mathcal{H}}_{0}(x_{i}%
)+\frac{1}{2}\sum_{j\neq i}V(x_{i},x_{j}).
\label{eq:2}%
\end{equation}
After performing the above referred minimization process by using the
Lagrange multipliers scheme, a \textquotedblleft fully unrestricted\textquotedblright\ set of $HF$ equations for the orbitals $\phi_{k_{i}}$ is obtained in the form

\[
\lbrack\hat{\mathcal{H}}_{0}(x)+\sum_{\eta_{1}}\sum_{s^{^{\prime}}}\int
d^{2}x^{^{\prime}}\phi_{\eta_{1}}^{\ast}(x^{^{\prime}},s^{^{\prime}%
})V(x,x^{^{\prime}})\phi_{\eta_{1}}(x^{^{\prime}},s^{^{\prime}})]\phi_{\eta
}(x,s)-
\]%
\begin{equation}
\sum_{\eta_{1}}[\sum_{s^{^{\prime}}}\int d^{2}x^{^{\prime}}\phi_{\eta_{1}%
}^{\ast}(x^{^{\prime}},s^{^{\prime}})V(x,x^{^{\prime}})\phi_{\eta}%
(x^{^{\prime}},s^{^{\prime}})]\phi_{\eta_{1}}(x,s)=\varepsilon_{\eta}%
\phi_{\eta}(x,s),\label{eq: 3 hartree}%
\end{equation}
where  $\eta,\eta_{1}=k_{1},...,k_{N}$. In this completely unrestricted
way, the $HF$ set of equations was wrote by Dirac \cite{Dirac}. In this form, the
orbitals are allowed to show arbitrary spin projection at any point of the
space, and this determines the use of the terms \textquotedblleft fully unrestricted\textquotedblright\ for this set of equations.
The total $HF$ energy of the $N$
electron system $E_{HF}$ and the $HF$ single particle energies  have the
forms

\begin{equation}
\begin{array}
[c]{ccc}%
E_{HF} & = & \sum_{\eta}\langle\eta|\hat{\mathcal{H}}_{0}|\eta\rangle+\frac
{1}{2}\sum_{\eta,\eta_{1}}\langle\eta,\eta_{1}|V|\eta_{1},\eta\rangle-\frac
{1}{2}\sum_{\eta,\eta_{1}}\langle\eta,\eta_{1}|V|\eta,\eta_{1}\rangle,\\
&  & \\
\varepsilon_{\eta} & = & \langle\eta|\hat{\mathcal{H}}_{0}|\eta\rangle
+\sum_{\eta_{1}}\langle\eta,\eta_{1}|V|\eta_{1},\eta\rangle-\sum_{\eta_{1}%
}\langle\eta,\eta_{1}|V|\eta,\eta_{1}\rangle.
\end{array}
\label{eq:4}%
\end{equation}
The definitions of the basis states of the one electron band model and their
scalar products appearing in the above equations are specified in Refs. \cite{Cabo1,Cabo2,Cabo3}.
\subsubsection{The $\alpha$ and  $\beta$ spin constraints.}

The analysis in Refs. \cite{Cabo1,Cabo2,Cabo3} was initially motivated by the aim of examining the
restrictions that could be introduced  in the description of many particle systems,
by the frequently employed assumption about that the $HF$ single particle states
should have a definite $+1/2$ ($\alpha)$ or $-1/2$ ($\beta$) projection of their spin at
all the spatial points. That is,  to show the structure:
\begin{equation}
\phi_{k}(x,s)=\left\{
\begin{array}
[c]{c}%
\phi_{k}^{\alpha}(x)u_{\uparrow}(s)\,\,\,\,type\,\,\alpha,\\
\phi_{k}^{\beta}(x)u_{\downarrow}(s)\,\,\,\,type\,\,\beta.
\end{array}
\right.  \label{eq:5}%
\end{equation}
Whenever the spatial functions  $\phi_{k}^{\alpha}(x)$ are identical, the $HF$
evaluation is called a $restricted$  one, and when they can be different it is
described as an  $unrestricted$ one \cite{Pickett}. However, both
of these cases are in fact restrictive ones for the allowed spin orientations
of the $HF$ orbitals. It  was emphasized in references \cite{Cabo1,Cabo2,Cabo3} that these
assumptions about the spin structure of the orbitals can be characterized as
definite constraints which drastically limits the generality of the space of
function in which the $HF$ single particle orbitals are
searched. It can be noted that the $HF$ scheme introduced by Dirac \cite{Dirac} does not
include any restriction on the spin structure of the searched orbitals.

\subsection{The model for the $CuO_{2}$ planes}

\begin{figure}[ht]
\par\begin{centering}
\includegraphics[scale=.45]{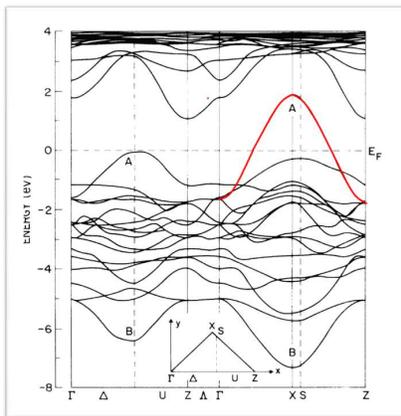}
\par\end{centering}
\caption{ Band structure of the  $La_{2}CuO_{4}$, calculated by Matheiss  1987. The single half-filled band
  suggests the consideration of the here investigated  model, which is based in a set of  interacting electrons
 subject to a crystalline $TB$ potential, tightly binding them to the points of the
  lattice formed by the $Cu$ atoms in the material. }%
 \label{Imagen4}
\end{figure}

\begin{figure}[ht]
\par\begin{centering}
{\scriptsize
(a)}\includegraphics[scale=0.3]{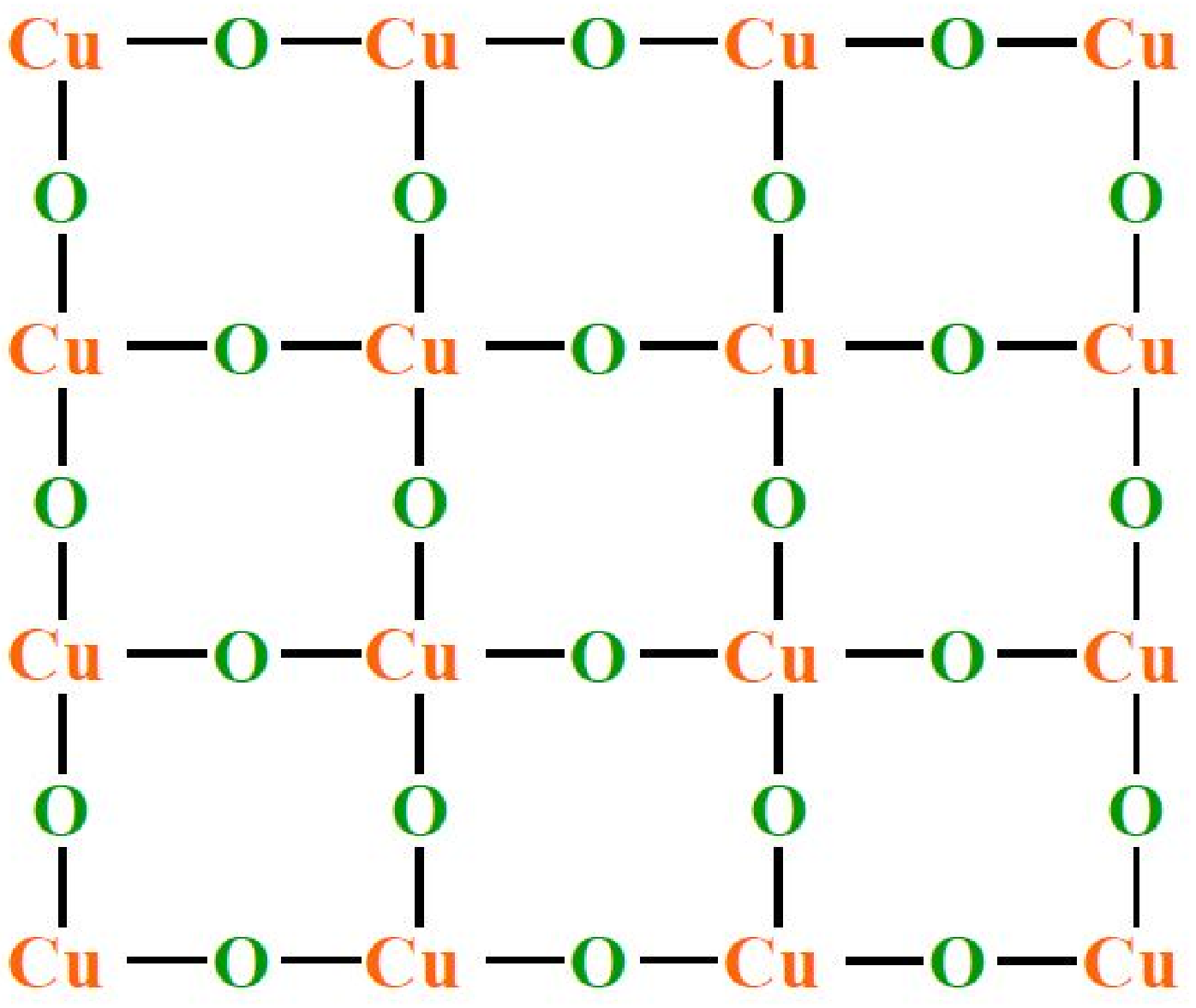}\hspace{8mm}{\scriptsize
(b)}~~\includegraphics[scale=0.35]{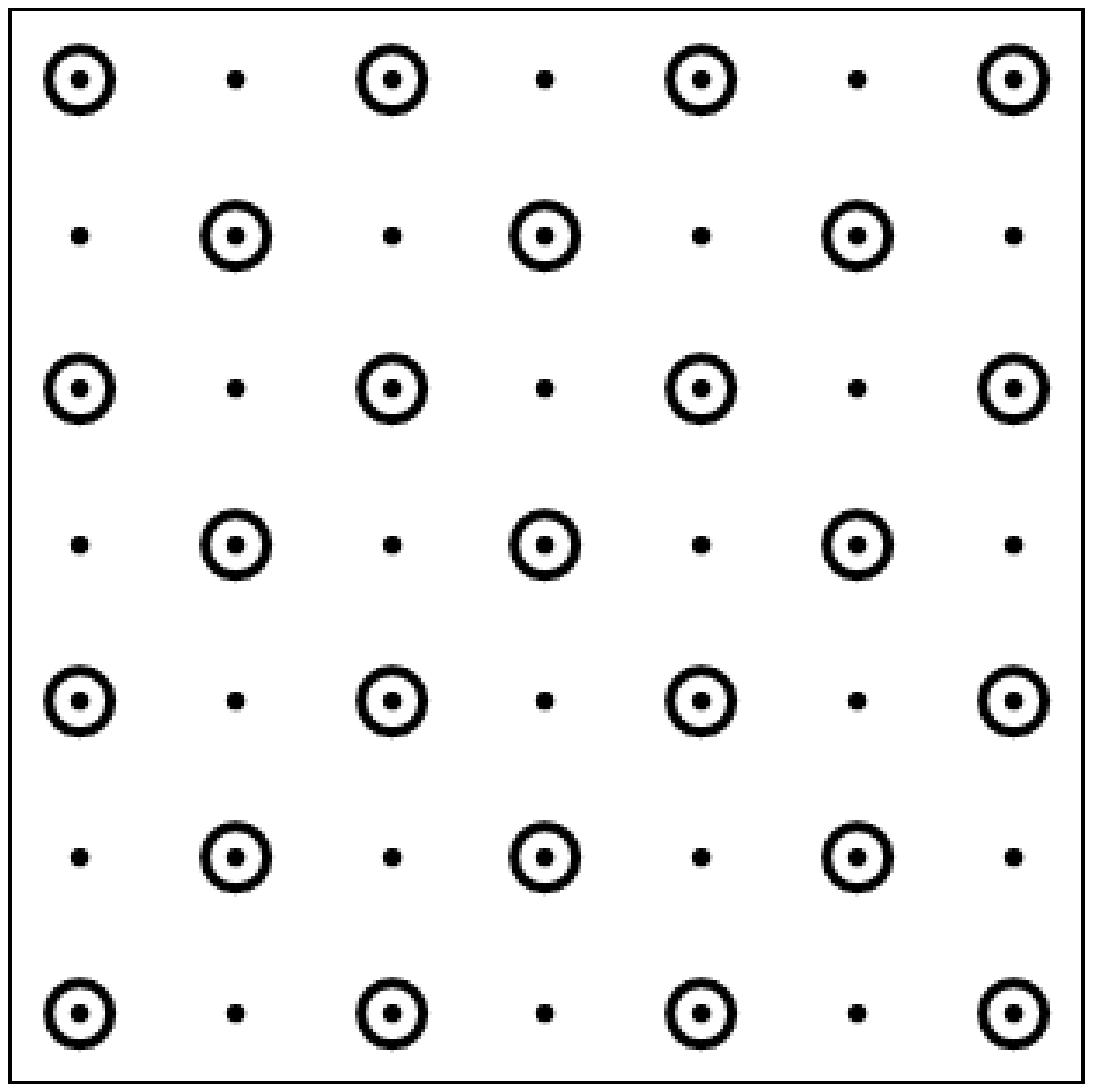}
\par\end{centering}
\caption{(a) The picture illustrates the $CuO_2$ planes  in the material.  (b)
Point lattice associated to those planes. In the search of the possibility of describing the $AF$ properties of the material through the elimination of the symmetry restrictions (over the
 space of functions in which the single particle states were  searched) it was helpful to decompose
 the  full lattice in the two sublattices which are differentiated  in the picture.}%
 \label{planocuoab}
\end{figure}

In figure \ref{Imagen4}, a calculated band diagram of the  $La_{2}CuO_{4}$ is illustrated. The
band structure was  obtained by the use of the $LAPW$ (Linear Augmented Plane
Waves) method \cite{Matheiss}. Let us describe below the formulation of
the model for the  $CuO_{2}$ planes introduced in Refs.
\cite{Cabo1,Cabo2,Cabo3}, which
is the starting point of the present work. It can be noted that the last
band occupied by electrons is half-filled, a fact that indicates a metallic
character for the material. The form of this band suggested the validity of an
approximate $TB$  description for the electrons populating it. The less bound kind of electron in  the compound  $La_{2}CuO_{4}$, is the unpaired one in the  $Cu^{2+}$, which at variance with the  $O^{2-}$ in the plane, does not has its last shell $(3d)$ closed. These electrons, in a
qualitative picture can be estimated to be the ones constituting the mentioned
unique band crossing the Fermi level in the band calculations done in the
Ref. \cite{Matheiss}. These electrons can be reasonably considered as
tightly  bound to the $Cu$ atoms in the $CuO_{2}$ planes, assumed the completion
of the  $O^{2-}$ atoms in those planes. The last remarks, support the idea
of taking as the lattice defining the $TB$ model, that one giving rise to the
half-filled band in Ref. \cite{Matheiss}. That is, the planar squared lattice
of points coinciding with the sites of the $Cu$ atoms in the $CuO_{2}$
plane (see figure \ref{planocuoab}).

The presence of all the other electrons filling the rest of the bands, in
common with all the neutralizing nuclear charges,   were  taken to play a double
role in the model defined  in Refs.
\cite{Cabo1,Cabo2,Cabo3}. In one sense, this system
is assumed as a polarizable effective medium which screens the Coulomb
repulsion  by means of effective dielectric constant  $\epsilon$. In
second place, this system is  assumed to generate a mean periodic Tight Binding
potential $W_{\gamma}$,  which strongly confines the electrons of  the single
partially filled band to be close to the $Cu$ atoms in the $CuO_{2}$ planes. The model  is assumed to be purely $2D$, that is, the space coordinates are
assumed to have only two planar components $\mathbf{x}=(x_{1},x_{2})$. It is
completed by assuming that the electrons in the partially filled \ band are
also interacting with the potential  $F_{b}$ generated by a jellium charge
density which neutralizes the net charge of the electrons of the model. The jellium charges were assumed as periodically concentrated and having a planar gaussian
distribution  with a radial width defined by the parameter $b$. In
resume, the free Hamiltonian for  the $TB$ model for the electrons in the half-filled band has the form
\begin{equation}
\hat{\mathcal{H}}_{0}(\mathbf{x})=\frac{\mathbf{p}^{2}}{2m}+W_{\gamma
}(\mathbf{x})+F_{b}(\mathbf{x}),\label{eq:6}%
\end{equation}
in which the periodic $TB$ potential is satisfying the periodicity condition%

\begin{equation}
W_{\gamma}(\mathbf{x})=W_{\gamma}(\mathbf{x}+\mathbf{R}),\label{eq:7}%
\end{equation}
and the jellium potential is defined by
\begin{equation}
F_{b}(\mathbf{x})=\frac{e^{2}}{4\pi\varepsilon_{0}\varepsilon}\sum
_{\mathbf{R}}\int d^{2}y\frac{\exp(-\frac{(\mathbf{y}-\mathbf{R})^{2}}{b^{2}%
})/\pi b^{2}}{\left\vert \mathbf{x}-\mathbf{y}\right\vert },\,\,\,b\ll
p,\label{eq:8}%
\end{equation}
in which the coordinate vectors of the planar $Cu$ atoms are
\begin{equation}\label{9}
\mathbf{R}=n_{x_{1}}p\,\mathbf{e}_{x_{1}}+ n_{x_{2}}p\,\mathbf{e}_{x_{2}}\qquad\mbox{where}\qquad
n_{x_{1}},n_{x_{2}}\in\mathbb{Z},
\end{equation}
where the unit vectors $\mathbf{e}_{x_{1}}$ and $\mathbf{e}_{x_{2}}$ lay on
directions defined by the vectors joining the nearest neighbors of the $Cu$ atoms in the lattice in figure
\ref{planocuoab}. It is known that the distance between a $Cu$ atoms and its nearest
neighbor is  $p=3.82$ ${\AA}$ \cite{Yanase}. Further, the Coulomb interaction
between two electrons in the partially filled band of the model is assumed in
the form
\begin{equation}
V=\frac{e^{2}}{4\pi\epsilon_{0}\epsilon}\frac{1}{\left\vert \mathbf{x}%
-\mathbf{y}\right\vert },\label{eq:10}%
\end{equation}
including a  dielectric constant which is supposed  to be generated by the
polarization of the set of electrons filling the other bands and all the nuclei in the
$La_{2}CuO_{4}$, through which the electrons of the partially filled band of
the model are assumed to move.

The model was started to be constructed with the initial idea of searching for $HF$ single particle
states of  the electrons in the partially filled band, not being of an
$\alpha$ or $\beta$ types, that is,  not being separable in their spatial and
spin dependence. In other words, allowing that  in those $HF$ orbitals the spin
projection could depend on the spatial position (spin-space entanglement or
non collinear spin structure). Materials showing $AF$ structure were suspected
to lead to $HF$ solutions of this kind, a fact that also suggested the
possibility of improving the understanding of the troublesome state of
knowledge of the band structure of such materials. In the particular
case of the $La_2CuO_4$, it is known that it is an antiferromagnetic compound. In the $AF$ systems, normally  the crystal symmetry of the substance  is directly broken by the $AF$ order. Therefore the
description of the model was also designed to incorporate the possibility of
the breaking of the crystalline translation invariance.
The $AF$ order has translation symmetry in each  one of the two interpenetrating squared sublattices in
which the planar $Cu$ lattices can be decomposed. Thus, the $HF$ single particle
orbitals in the model were assumed to retain the Bloch functions character
under the common reduced translation symmetry group of these sublattices. Then,
the physical state should be equivalent under the translations leaving
invariant the sublattices, but not under the one which transforms a sublattice
into the other. This reduced symmetry is associated to a subgroup of the
total translation group of the crystal and therefore its space representation
should be more reduced in its number of states as classified by the momenta
$\mathbf{k}$ in the reciprocal space.

The two sublattices of point depicted in figure \ref{planocuoab}, that will be described below by the
indices $r=1$ or $2$, are analytically defined in the form
\begin{equation}
\mathbf{R}^{(r)}=\sqrt{2}n_{1}p\,\mathbf{q}_{1}+\sqrt{2}n_{2}p\,\mathbf{q}%
_{2}+\mathbf{q}^{(r)},\label{eq:11}%
\end{equation}%
\[
\mathbf{q}^{(r)}=\left\{
\begin{array}
[c]{c}%
\mathbf{0}\,\,\,\,\,\,\,\,\,if\,\,r=1\\
p\,\mathbf{e}_{x_{1}}\,\,if\,\,r=2,
\end{array}
\right.
\]
where  $\mathbf{q}_{1}$ and $\mathbf{q}_{2}$ are the two unit vectors defining
the directions of the unit cell vector of the sublattices.

\subsubsection{Sublattice translations }

\begin{figure}[ht]
\begin{centering}
\includegraphics[scale=0.3]{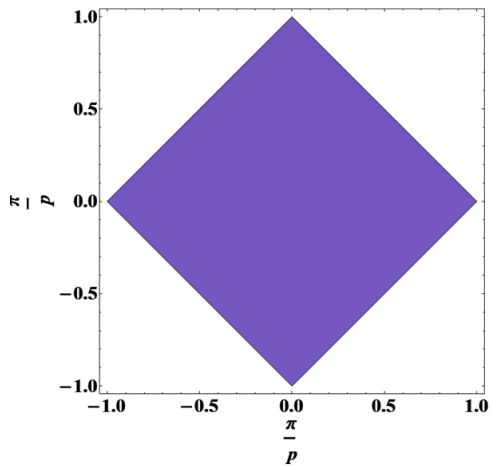}
\par
\end{centering}
\caption{ The darker region shows the reduced Brillouin zone associated to the Bloch functions
defined on the sublattices. The length of its sides is $\sqrt{2}\frac{\pi}{p}$.}%
\label{brillouin}
\end{figure}

Then, the searched $HF$ single particle wave functions should be eigenfunctions
of the discrete translation operator $\hat{T}_{\mathbf{R}^{(1)}}$, which
transform a sublattice in itself
\begin{equation}
\hat{T}_{\mathbf{R}^{(1)}}\phi_{\mathbf{k},l}=\exp(i\mathbf{k}\cdot
\mathbf{R}^{(1)})\phi_{\mathbf{k},l}.\label{eq:12}%
\end{equation}
When the point lattice is infinite, the  Brillouin cell formed by the set of all the
momenta $\mathbf{k}$ indexing the eigenfunctions of  $\hat{T}%
_{\mathbf{R}^{(1)}}$ is the shadowed zone in figure \ref{brillouin}, while the large square
represents the set of momenta associated to the eigenfunctions of the  group
of translations in the original lattice. Since it is impossible to numerically
treat the infinite lattices because it has a continuum of states, periodical
boundary conditions were chosen in order to make finite the number of
eigenfunctions in the Brillouin zone. The boundary conditions were imposed
fixing the periodicity of the eigenfunctions $\phi_{\mathbf{k},l}$ in the
boundaries  defined by  $\ x_{1}=-Lp$ and $x_{1}=Lp$ along the $x_{1}$ axis and
$\ $by $\ x_{2}=-Lp$ and $x_{2}$= $Lp$ along the $x_{2}$ axis. The
finite set of momenta vectors obeying the boundary conditions are
\[
\mathbf{k}=\left\{
\begin{array}
[c]{c}%
\frac{2\pi}{Lp}(n_{x_{1}}\mathbf{e}_{x_{1}}+n_{x_{2}}\mathbf{e}_{x_{2}})\\
n_{x_{1}},n_{x_{2}}\in\mathbb{Z}\\
-\frac{L}{2}\leq n_{x_{1}}\pm n_{x_{2}}<\frac{L}{2}.
\end{array}
\right.
\]
The number of electron eigenfunctions obeying the conditions is  $\frac{N}{2}$, that is,  a half of the number of functions satisfying the same boundary periodicity conditions when the group is the group of translations in the full point lattice.

\subsubsection{Tight-Binding single band basis}

The $TB$ basis  that was  employed for describing the one band model had the
form

\begin{equation}
\varphi_{\mathbf{k}}^{(r,\sigma_{z})}(\mathbf{x},s)=\sqrt{\frac{2}{N}%
}u^{\sigma_{z}}(s)\sum_{\mathbf{R}^{(r)}}\exp(i\mathbf{k}\cdot\mathbf{R}%
^{(r)})\varphi_{0}(\mathbf{x}-\mathbf{R}^{(r)}),\label{eq:13}%
\end{equation}%
\[
\hat{\sigma}_{z}u^{\sigma_{z}}=\sigma_{z}u^{\sigma_{z}},
\]%
\begin{equation}
\varphi_{0}(\mathbf{x})=\frac{1}{\sqrt{\pi a^{2}}}\exp(-\frac{\mathbf{x}^{2}%
}{2a^{2}}),\,\,\,a\ll p.\label{eq:14}%
\end{equation}
where  $N$ is the number of electrons in the model electron gas filling
part of the band crossing the Fermi level and  $\hat{\sigma}_{z}$ is the spin
projection operator in the  $z$ axis, which is assumed to be orthogonal to
the $CuO_{2}$ planes ; $\sigma_{z}=-1\,,\,\,1$ are its
eigenvalues; $r=1\,\,$and$\,\ 2$, is the above defined index
for each of the two sublattices. Note that in the small overlapping
approximation between nearest neighborhood points (which should be in different
sublattices), the orthogonality between the members of this basis is only
lost between functions belonging to different sublattices  having the same
spin polarization and momentum. The orthogonality between distinct elements
corresponding to the same sublattice is valid by construction.

The chosen Wannier orbital of the model  $\varphi_{0}(\mathbf{x}%
-\mathbf{R}^{(r)})$ were fixed for simplicity. Their gaussian form implies that it was supposed that the net $TB$ potential in the close neighborhood of
each $Cu$ atom had been assumed to be a simple 2D harmonic potential. This
simplifying assumption  was adopted following the idea that the main forces in defining the $AF$ and insulator properties of the $CuO_2$ planes were in fact determined by the spontaneous
breaking of the crystalline symmetry plus the spin entangled structure of the
$HF$ electron orbitals. Thus, simplifying  considerations led to fix the  explicit form of equation (\ref{eq:13}) for the starting basis states of the model. However, taking into account  improved definitions  for the Wannier's orbitals defining the single band model,  might be of  help for the description of other effects, by example  the  magnetic anisotropy of the $AF$ order in $La_2CuO_4$. For this purpose, the model  could be generalized by considering atomic $3D$ representation, employing  Wannier's orbitals like the $d$  ones of the  $Cu$ and the inclusion of spin-orbit interactions. This study is expected to be considered elsewhere.
\subsection{Hartree-Fock solution with hole doping}

In this subsection we present  the matrix problem to which the $HF$ set of
equations of the model was reduced in Refs.
\cite{Cabo1,Cabo2,Cabo3}, after its
projection on the defined basis. In the present work, these equations will
be solved for the more general situation in which the system is doped with
holes. In the coming subsection it will be  reviewed  how the model reproduces the main characteristics
of the dispersion of the single band crossing the Fermi level in Ref. \cite{Matheiss}
(predicting a metallic and paramagnetic state) when full crystalline
symmetry and the $HF$ orbitals of the type $\alpha$ and $\beta$ are assumed.
Afterwards, the next sections will illustrate the
consequences of the elimination of the symmetry restrictions in the space of
functions in which the $HF$ energy functional is minimized. It was performed in a similar way as
it was done in Refs.
\cite{Cabo1,Cabo2,Cabo3}, but for the important situation in
which the systems is doped with holes.  At very small doping values,  the solutions
are the ones obtained in Refs.
\cite{Cabo1,Cabo2,Cabo3} for the  exact half-filling situation: the ground state is an antiferromagnetic-insulator and the excited phase corresponds to the
paramagnetic-pseudogap state. When the hole doping is augmented the
results indicate the appearance of a quantum phase transition at the doping
value  $x=0.2$, in which the $AFI$ and $PPG$ states both coalesce in one single
metallic state for higher doping values.

As in \ Refs.
\cite{Cabo1,Cabo2,Cabo3} the searched $HF$ single particle states are expressed as
a linear combination of the before defined  basis functions  in the form
\begin{equation}
\phi_{\mathbf{k},l}(\mathbf{x},s)=\sum_{r,\sigma_{z}}B_{r,\sigma_{z}%
}^{\mathbf{k},l}\varphi_{\mathbf{k}}^{(r,\sigma_{z})}(\mathbf{x}%
,s),\label{eq:15}%
\end{equation}
where $l$ is the index of the quantum numbers required to uniquely define the $HF$ single particle states. After substituting  the above
expression for the orbitals in the $HF$ equations and taking their scalar product with an arbitrary state, the equivalent self-consistent matrix problem can be written in the form
\begin{equation}
\left[  E_{\mathbf{k}}^{0}+\tilde{\chi}(G_{\mathbf{k}}^{Co}-G_{\mathbf{k}%
}^{int}-F_{\mathbf{k}})\right]  \cdot B^{\mathbf{k},l}=\tilde{\varepsilon}%
_{l}(\mathbf{k})I_{\mathbf{k}}\cdot B^{\mathbf{k},l},\label{eq:16}%
\end{equation}
with the definitions for the constants
\begin{equation}
\tilde{\chi}=\frac{me^{2}a^{2}}{4\pi\hbar^{2}\epsilon\epsilon_{0}%
p},\label{eq:17}%
\end{equation}%
\begin{equation}
\tilde{\varepsilon}_{l}(\mathbf{k})=\frac{ma^{2}}{\hbar^{2}}\varepsilon
_{l}(\mathbf{k}),\label{eq:18}%
\end{equation}
which are dimensionless, as also are all the implicit parameters in the definitions
of all the entering  matrices
\[
E_{\mathbf{k}}^{0}=\left\Vert E_{\mathbf{k},(t,r,\alpha_{z},\sigma_{z})}%
^{0}\right\Vert _{4\times4}\,,
\]%
\[
G_{\mathbf{k}}^{C}=\left\Vert G_{\mathbf{k},(t,r,\alpha_{z},\sigma_{z})}%
^{C}\right\Vert _{4\times4}\,,
\]%
\[
G_{\mathbf{k}}^{i}=\left\Vert G_{\mathbf{k},(t,r,\alpha_{z},\sigma_{z})}%
^{i}\right\Vert _{4\times4}\,,
\]%
\[
F_{\boldsymbol{k}}=\left\Vert F_{\mathbf{k},(t,r,\alpha_{z},\sigma_{z}%
)}\right\Vert _{4\times4}\,,
\]%
\[
I_{\boldsymbol{k}}=\left\Vert I_{\mathbf{k},(t,r,\alpha_{z},\sigma_{z}%
)}\right\Vert _{4\times4}\,.
\]
They are respectively associated to the kinetic term plus the periodic $TB$
potential, the Coulomb direct and exchange potential terms, the potential
generated by the compensating jellium charge density and the overlapping
matrix between the basis functions. The formulae for  the matrix elements
are shown in the Appendices of the Refs.
\cite{Cabo2,Cabo3}. In the
employed representation the normalization condition of the $HF$ single particle
states and the formula for the $HF$ energy take the forms%
\begin{eqnarray}
1 & = & B^{\mathbf{k},l^{\ast}}.I_{\boldsymbol{k}}.B^{\mathbf{k},l},\\
E_{\mathbf{k},l}^{HF} & = & \sum_{\mathbf{k},l}\Theta_{(\tilde{\varepsilon
}_{F}-\tilde{\varepsilon}_{l}(\mathbf{k}))}[\tilde{\varepsilon}_{l}%
(\mathbf{k})-\frac{\tilde{\chi}}{2}B^{\mathbf{k},l\ast}.(G_{\mathbf{k}}%
^{C}-G_{\mathbf{k}}^{i}).B^{\mathbf{k},l}].
\label{eq:19}%
\end{eqnarray}

The $HF$ matrix system of equations (\ref{eq:16}) is a non linear one in the variables
$B_{r,\sigma_{z}}^{\mathbf{k},l}$, which are the four  components of the
vector $\ B^{\mathbf{k},l}$ for each value of $\mathbf{k}$. The $B$ constants
can be interpreted  as determining the probability amplitudes to find the electron
in the states  $(\mathbf{k},l)$, of the sublattice $r$, and spin along the
$z$ axis. Note that for each  $\mathbf{k}$ \ value, four values of  the quantum numbers of the $HF$ single particle states  $l=1,2,3,4$ should be obtained, or what is equivalent, four bands in the reduced $BZ$.
From the equations (\ref{eq:16}) it  can be observed that the full matrix
representation of the  Fock operator is block diagonal in the momenta
indices  $\mathbf{k}$, which is a direct consequence of the symmetry with respect to the reduced translation invariance of the system.

\subsection{Maximal symmetry solutions: fixing the model parameters  }

\begin{figure}[ht]
\begin{centering}
\includegraphics[scale=0.25]{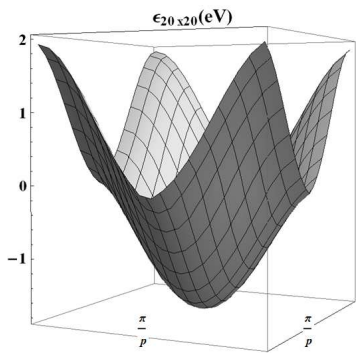}
\par\end{centering}
\caption{ The doubly degenerated paramagnetic and metallic band arising from the  $HF$ solution with full translation
symmetry and $\alpha$ and $\beta$ spin structure of the single particle states.}%
\label{parametalico}
\end{figure}

In first place the $HF$ problem was solved by assuming that the $HF$ single particle states satisfied the translation invariance in the whole lattice and also were showing $\alpha$ or $\beta$ spin dependence. After this assumptions, as it was suspected, the $HF$ solution of the model produced the single particle spectrum depicted in the figure \ref{parametalico}, in a similar way as in Refs. \cite{Cabo1,Cabo2,Cabo3}. As before  we will adjust the free parameters $\epsilon$: the dielectric constant of the effective medium; $\tilde{a}$: the radial distance
for which the gaussian Wannier's orbitals are appreciably different from zero; $\tilde{\gamma}$: the hopping probability between nearest neighbor sites fixed by the effective medium and  $\tilde{b}$: radial distances inside which the jellium charges are concentrated.

The  Bloch Tight-Binding single band basis for this problem had the form
\begin{equation}
\varphi_{\mathbf{Q}}^{\sigma_{z}}(\mathbf{x},s)=\sqrt{\frac{1}{N}}%
u^{\sigma_{z}}(s)\sum_{\mathbf{R}}\exp(i\mathbf{Q}\cdot\mathbf{R})\varphi
_{0}(\mathbf{x}-\mathbf{R}),\label{eq:20}%
\end{equation}
where the appearing momenta $\mathbf{Q}$ \ are given  by
\[
\mathbf{Q}=\left\{
\begin{array}
[c]{c}%
\frac{2\pi}{Lp}(n_{x_{1}}\mathbf{e}_{x_{1}}+n_{x_{2}}\mathbf{e}_{x_{2}})\\
n_{x_{1}},n_{x_{2}}\in\mathbb{Z}\\
-\frac{L}{2}\leq n_{x_{1}},n_{x_{2}}<\frac{L}{2}%
\end{array}
\right.
\]
These functions define  the Bloch states  forming the basis of the maximal
translation group with periodic boundary conditions in the same region defined
before. In addition  $N=L\times L$, and  $\mathbf{R}$ are the number of cells
in the total lattice of points and the vectors defining the positions of the
lattice.  Let the $HF$ single particle states expressed in the form
\begin{equation}
\phi_{\mathbf{Q},l}(\mathbf{x},s)=\sum_{\sigma_{z}}B_{\sigma_{z}}%
^{\mathbf{Q},l}\varphi_{\mathbf{Q}}^{\sigma_{z}}(\mathbf{x},s),\label{eq:21}%
\end{equation}
in the just defined basis. In this case, the equivalent $HF$ matrix problem
obtained after substituting these functions in the set of $HF$ equations, is now
of dimension two. That is,  there are two components $l=1,2$ $\ $for
each\textbf{ }momentum $Q$. In this way, in an analogous form as in
(\ref{eq:16}) is possible to obtain the $HF$ matrix problem in the form
\begin{equation}
\left[  E_{\mathbf{Q}}^{0}+\tilde{\chi}(G_{\mathbf{Q}}^{C}-G_{\mathbf{Q}}%
^{i}-F_{\boldsymbol{Q}})\right]  \cdot B^{\mathbf{Q},l}=\tilde{\varepsilon
}_{l}(\mathbf{Q})I_{\mathbf{Q}}\cdot B^{\mathbf{Q},l},\label{eq:22}%
\end{equation}
which is a system of non linear matrix equations to be solved by iterations.

For starting the iterative process an initial paramagnetic state is employed
as an ansatz.  The figure  \ref{parametalico} shows the doubly degenerated metallic and
paramagnetic band which is obtained at half-filling  conditions. That is,
with  $N=20\times20$ electrons. The chosen parameters were: $\epsilon=12.5$,
which is a typical value in semiconductor systems. In what follows we will
explain the reasons for selecting the values  $\tilde{a}=0.25$, $\tilde
{b}=0.05$ \ and  $\tilde{\gamma}=0.03$, always following the criterium of
obtaining a band width of  $3.8\,eV$ as it is suggested by the band diagram in
figure \ref{Imagen4}.

Observe the coincidence in form between the calculated band and the
conduction one appearing in figure \ref{Imagen4}. In both of them, the Fermi level is
crossed by the dispersion curve at the mid point between the top and the
bottom of the bands, in the direction  $\Gamma$-$X$, while in the direction
laying at $45^{o}$ respect to $\Gamma$-$X$ the Fermi level becomes tangent to
the dispersion curve in its maximum.\bigskip

\section{Evolution of the paramagnetic-pseudogap band spectrum }

In this section we will present the results of solving the $HF$ matrix set of
equations for a variable hole doping in the case of the excited solution which
evolves from the pseudogap state obtained in Refs.
\cite{Cabo1,Cabo2,Cabo3} for the  exact
half-filling situation. It is important to recall that this state emerges
as a solution after only eliminating the restriction on the single particle
states of being a Bloch function in the full lattice formed by all the  planar $Cu$
atoms within the $CuO_{2}$ planes. The constraints of being single particle
states of $\alpha$ or $\beta$ types were yet maintained. This process allows to obtaining of a doubly degenerated $HF$ solution which shows a  pseudogap resembling the one observed in the normal state of the $HTc$ superconductor materials.

\begin{figure}[ht]
\begin{centering}
{\scriptsize
(a)}\includegraphics[scale=0.35]{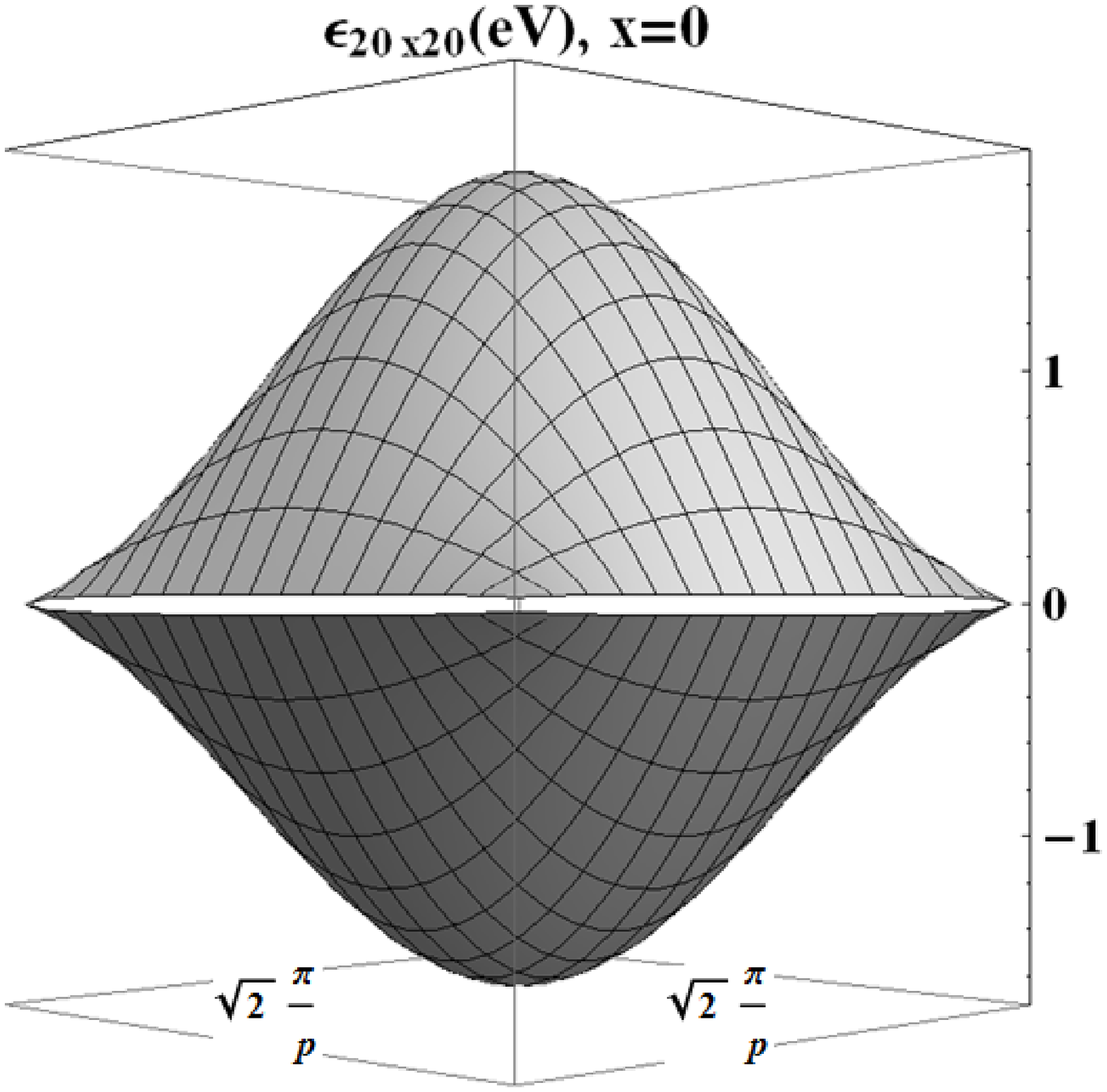}{\scriptsize (b)}
\includegraphics[scale=0.35]{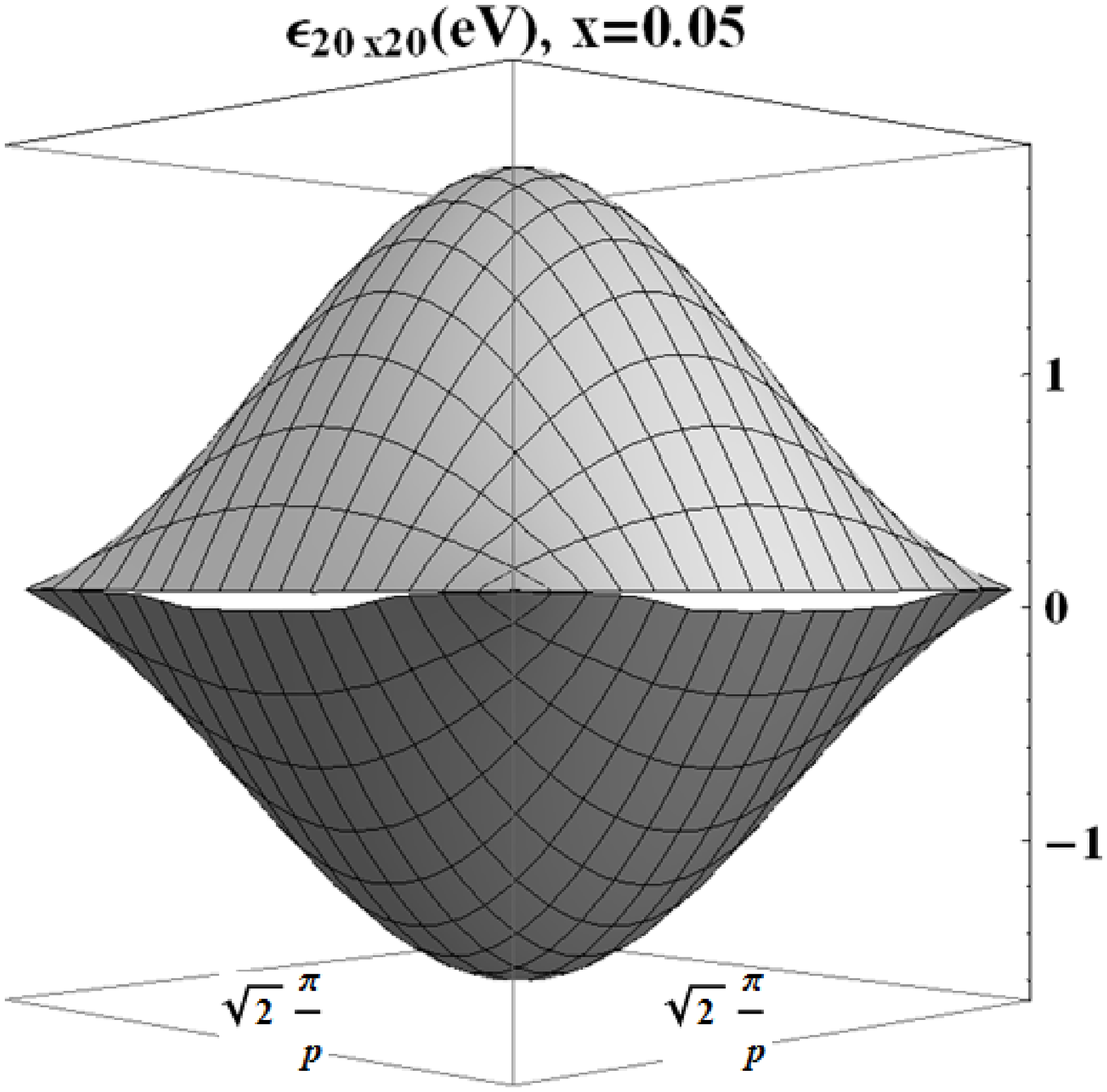}
\par\end{centering}
\par
\begin{centering}
\vspace{3mm}
\par\end{centering}
\par
\begin{centering}
{\scriptsize
(c)}\includegraphics[scale=0.35]{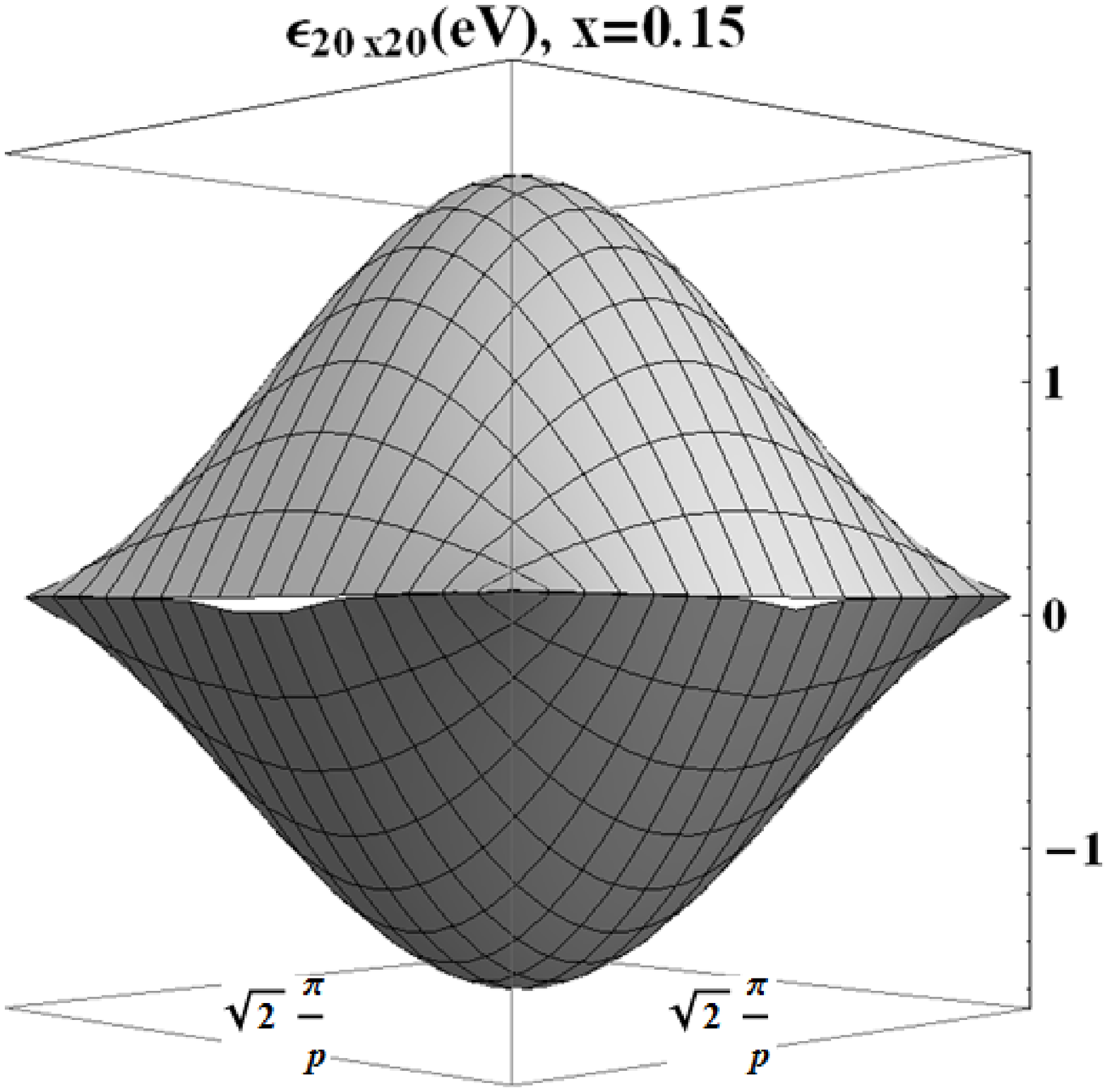}{\scriptsize
(d)}
\includegraphics[scale=0.35]{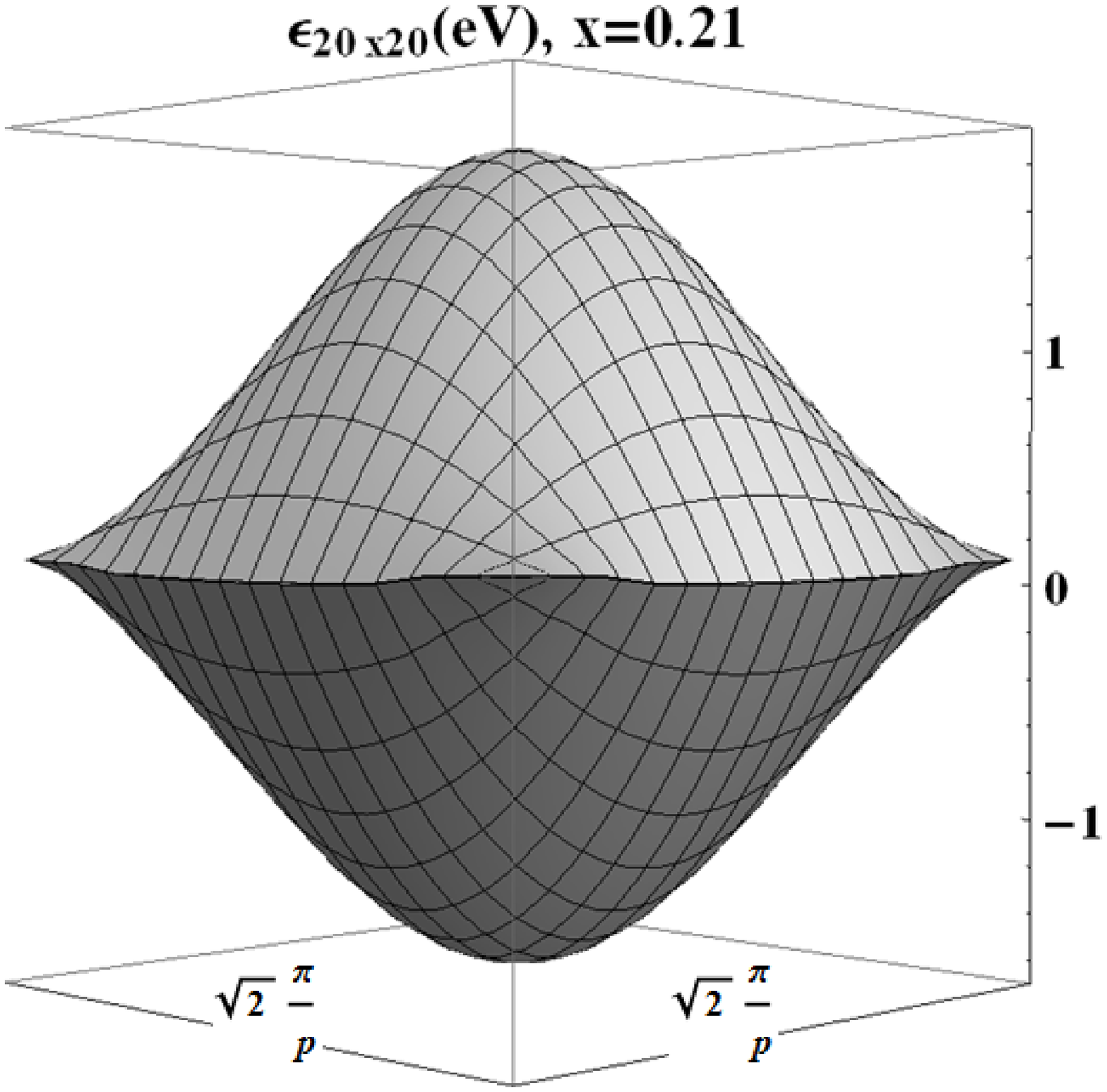}
\par\end{centering}
\caption{The evolution of the band spectrum of the $PPG$ as the hole concentration grows. The Fermi level lays in the zero of the graphics.  It can   be observed how the pseudogap starts closing at the corners of the Brillouin zone as the doping increases. }%
\label{dopingppg}
\end{figure}

Figure \ref{dopingppg} shows the evolution with increasing hole concentration of the
bands of the  pseudogap state, obtained for a point lattice of  $20\times20$
points. All the graphics are plotted in the Brillouin zone of the sublattices,
that is the darker zone in the figure \ref{brillouin}. Note the existence of a
pseudogap  which attains its maximum of the order $\sim80\,\,meV$. The
parameters fixed in the previous section were employed for this evaluation. The
maximal value for the gap appears at the mid points of the sides of the
Brillouin zone of the sublattices and furnishes an estimate of  $T\simeq
1000\,K$ for the temperature at which the pseudogap starts to be observed in
the experiments \cite{Tallon,loram,wuyts,Fauque}.

The evaluated result for the pseudogap depends from the
effective dielectric constant $\epsilon$ and the set of parameters which
were fixed in order to reproduce the single band crossing the Fermi level in
the band calculation done by Matheiss \cite{Matheiss}. Then, the value
obtained here should not be taken as precise prediction for the temperature of
the observed pseudogap. However, the ARPES experimental results for the
doped  $La_{2}CuO_{4}$ indicates a pseudogap temperature in the region
$100$-$200\,meV$, that is value close to $1000\,K$. Thus, the  $PPG$
Hartree-Fock solution offers a reasonable estimate of the pseudogap
temperature $T^{\ast}$ \cite{Tallon,loram,wuyts,Fauque}.

One interesting results which can be observed from figure \ref{dopingppg} is that by
augmenting the hole concentration, the so-called \textquotedblleft pockets\textquotedblright\ \cite{Plate,Shen,Doiron} in the corners of the Brillouin zone are formed. In addition it can observed that the pseudogap starts to
diminish  first in the corners until it fully collapses at the critical
doping  $x_{c}=0.2$. After this critical concentration of holes, the state
starts behaving as a paramagnetic metal.\bigskip

\section{Evolution of the antiferromagnetic-insulator band spectrum }

In this section we will consider the solution of the $HF$ system of equations
(\ref{eq:16}) by again using the successive iterations method, but in this case,
starting from an antiferromagnetic initial ansatz for the Slater determinant
state. The results  presented below, were found for the
values of the  set of  free parameters: $\epsilon=12.5$,
$\tilde{a}=0.25$, $\tilde{\gamma}=0.03$ and \textbf{$\tilde{b}=0.05$}, which were
also employed in the previous section.

\begin{figure}[ht]
\begin{centering}
{\scriptsize
(a)}\includegraphics[scale=0.35]{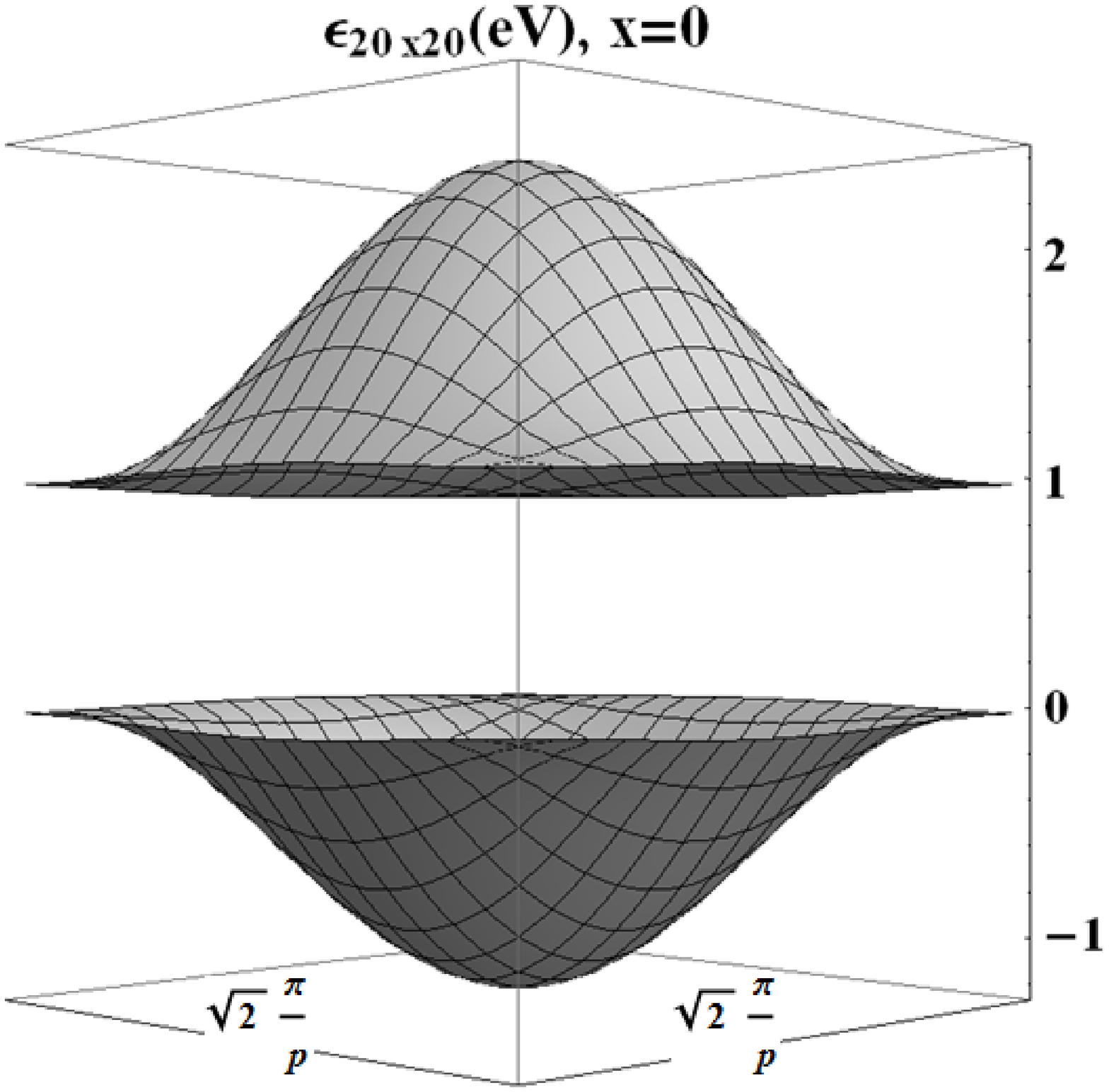}{\scriptsize
(b)}\includegraphics[scale=0.35]{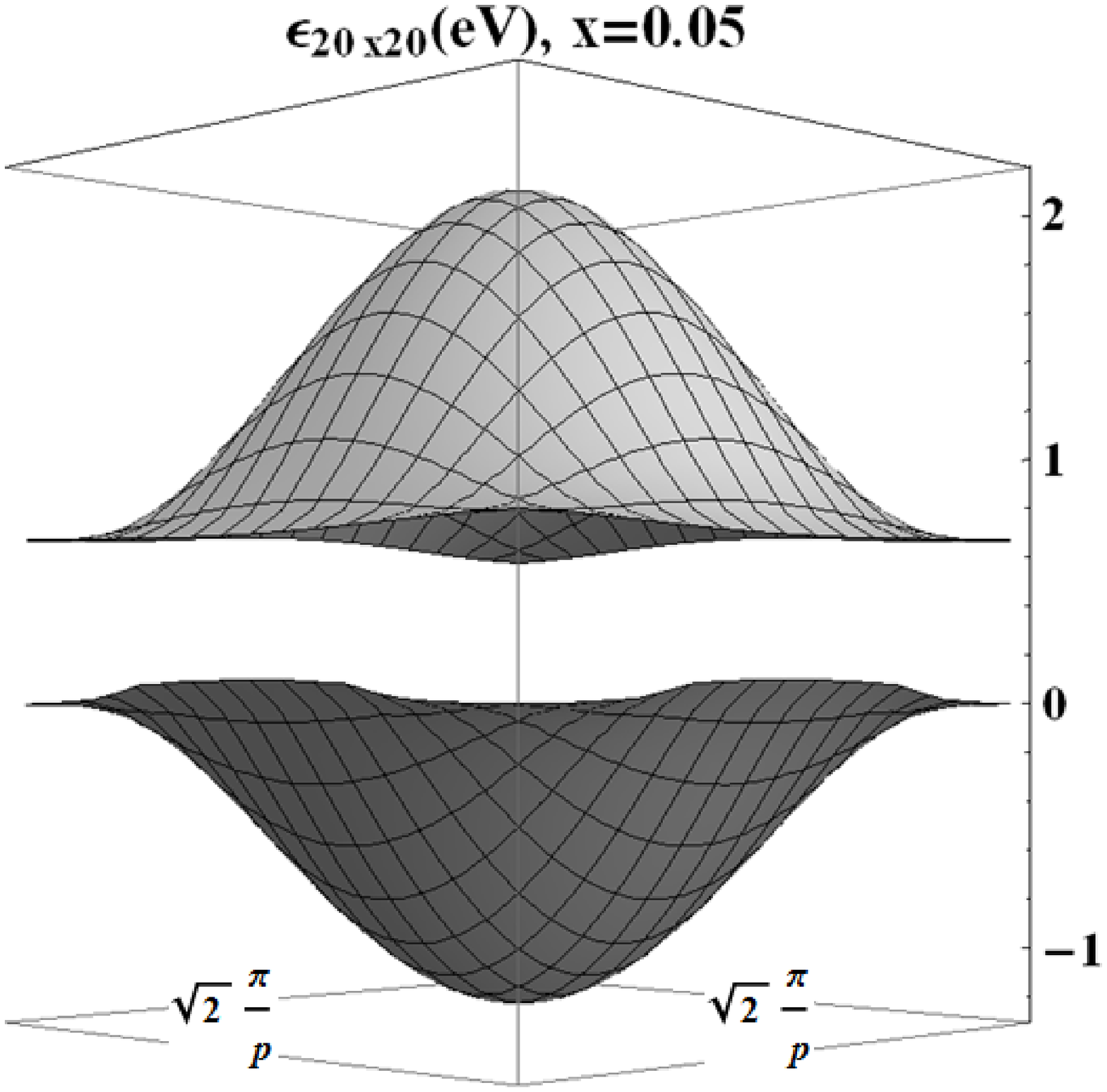}
\par\end{centering}
\par
\begin{centering}
\vspace{3mm}
\par\end{centering}
\par
\begin{centering}
{\scriptsize
(c)}\includegraphics[scale=0.35]{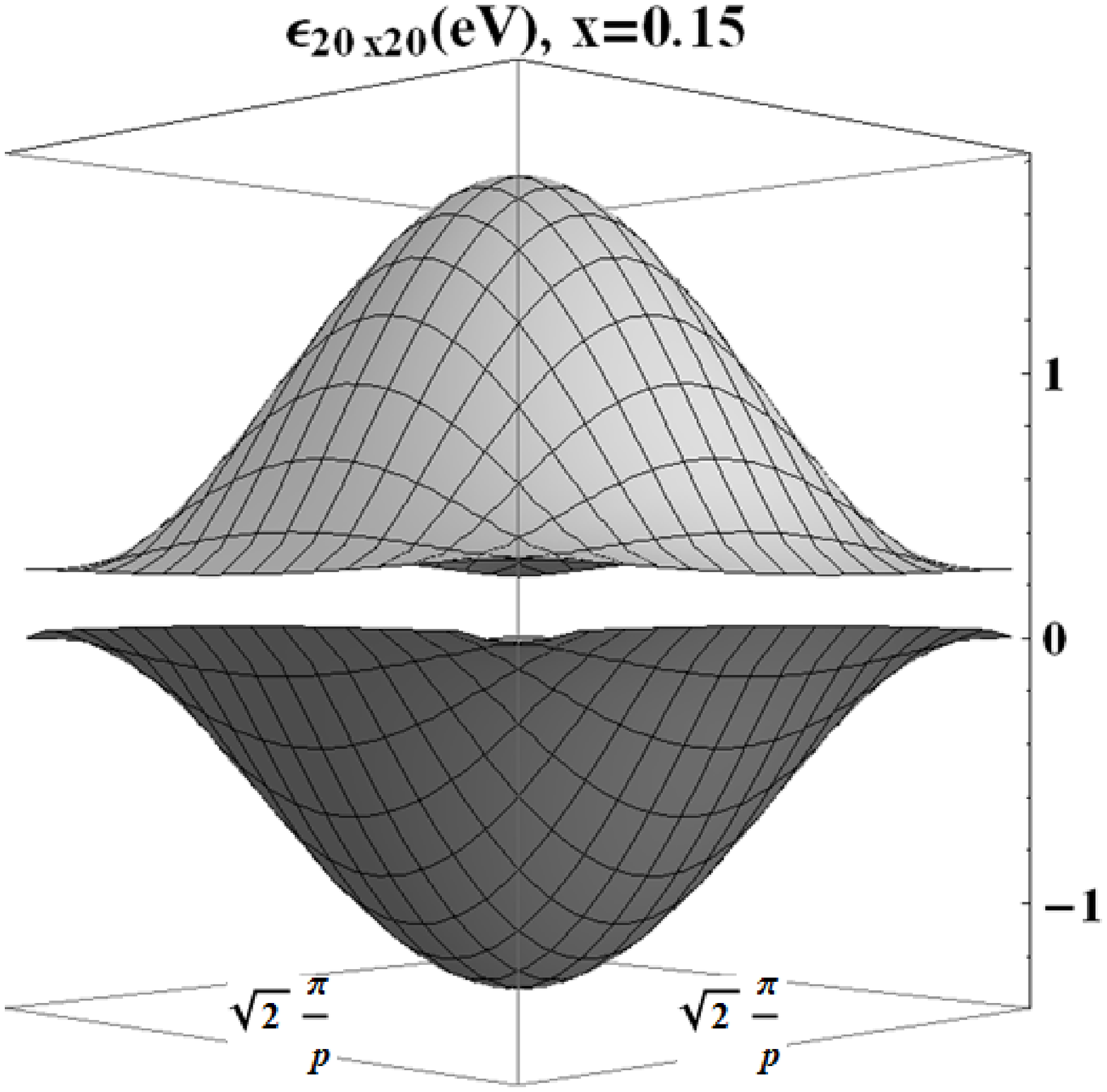}{\scriptsize
(d)}\includegraphics[scale=0.35]{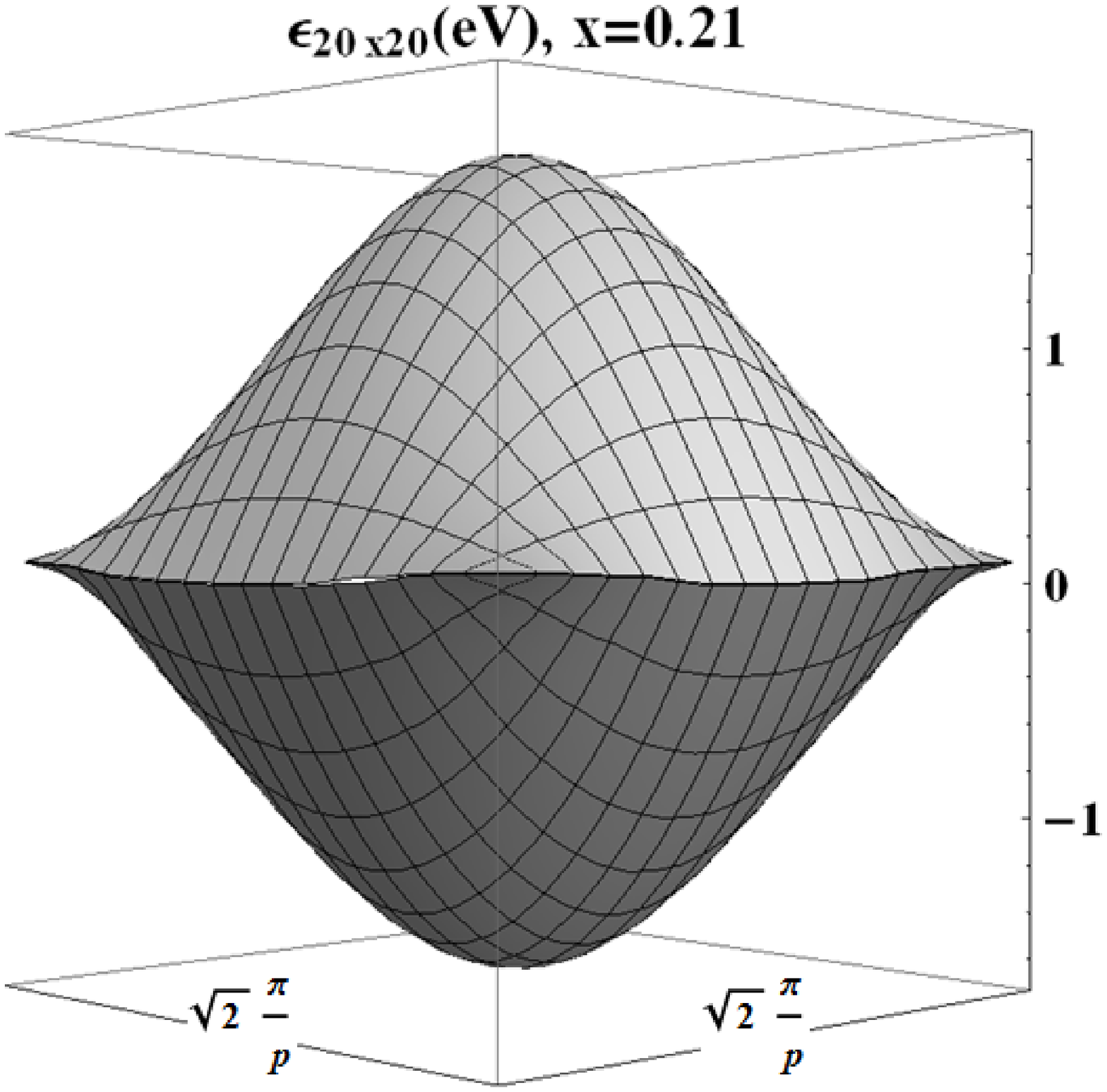}
\par\end{centering}
\caption{ Evolution  of the band spectrum $AFI$ as the hole doping increases.
 The Fermi level is laying in the zero of the plots. The hole states form the
the so-called Fermi arcs at the middle of the  sides of the reduced Brillouin cell and the gap diminishes until it is fully
closed  at the critical doping  $x_{c}=0.2$. }%
\label{dopingafa}
\end{figure}

The figure \ref{dopingafa} (a) shows the  band obtained for the
point lattice of  $20\times20$ points. The bands are clearly
associated to an insulator system, which resulted as the ground
state with respect to the pseudogap phase.

It also shows a local magnetization. These results are
essentially the same as the ones obtained in Refs. \cite{Cabo1,Cabo2,Cabo3} but for slightly different values for the free parameters of the model. The next figures \ref{dopingafa} (b), (c), (d)  show the evolution with the increasing hole doping of the band spectrum of these  states.
It can be observed that the added holes start accumulating at the mid points of
the reduced Brillouin zone. That is,  in the more energetic electronic
states, and as a consequence,  more preferable  by the holes.  All the
pictures are plotted in the reduced Brillouin zone of the
sublattices. When the doping is further increased up to some
critical point, the holes, which were sitting at the mid sides of the
Brillouin zone, drastically pass to occupy the corners of the
reduced Brillouin zone. This constitute a structural change of
the band occurring at the critical value of the hole doping
$\delta_{c}=0.2$, in which the form of the bands becomes different
before and after the states passes through this particular hole
concentration. In other words a quantum critical phase transition
occurs at this value of the hole doping. Thus, the model
predicts a phase transition which had being argued to exist
beneath the superconductor dome \cite{Broun,Sachdev}.

\begin{figure}[ht]
\par\begin{centering}
{\small (a)}\includegraphics[scale=0.5]{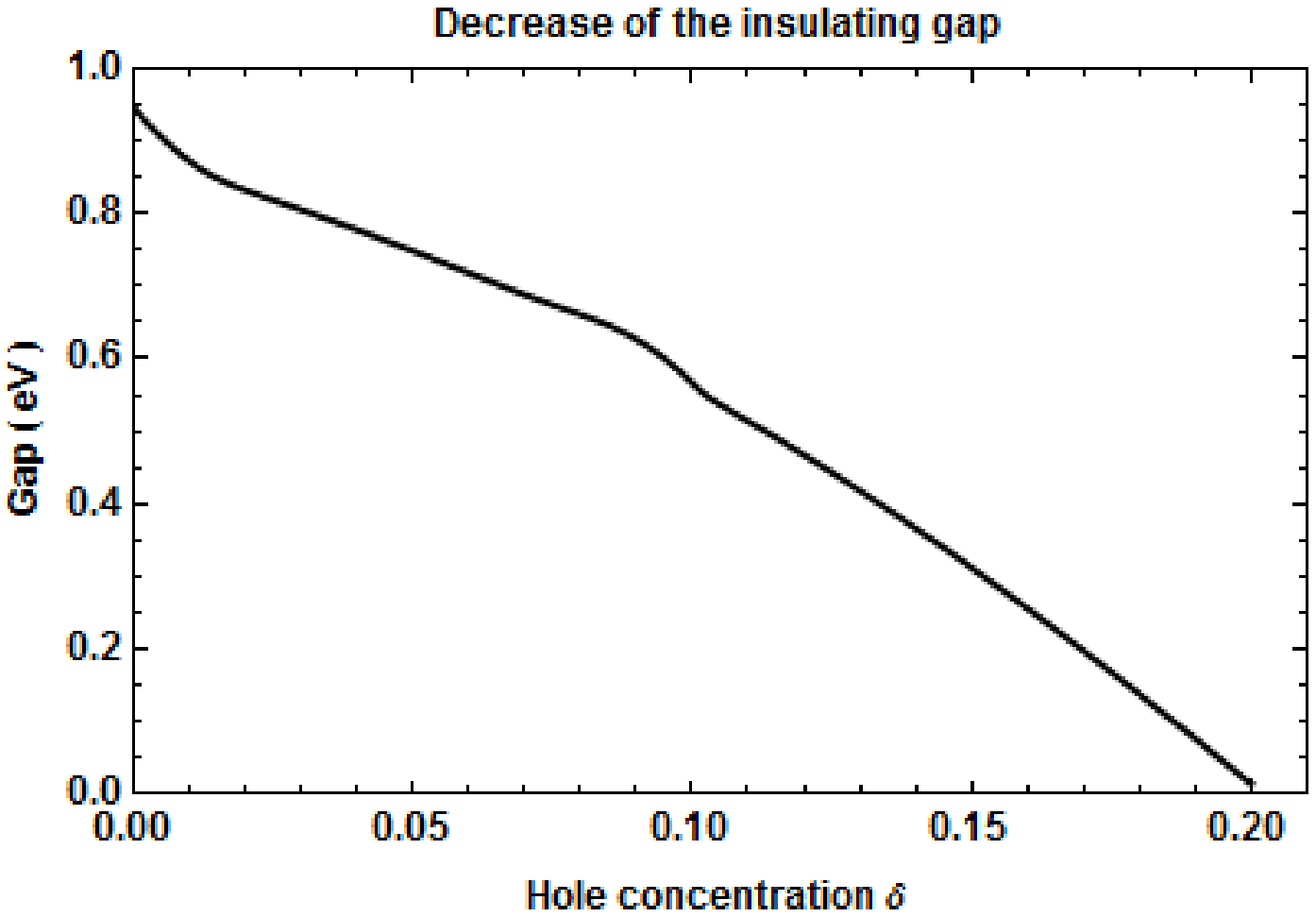}{\small
(b)}\includegraphics[scale=0.5]{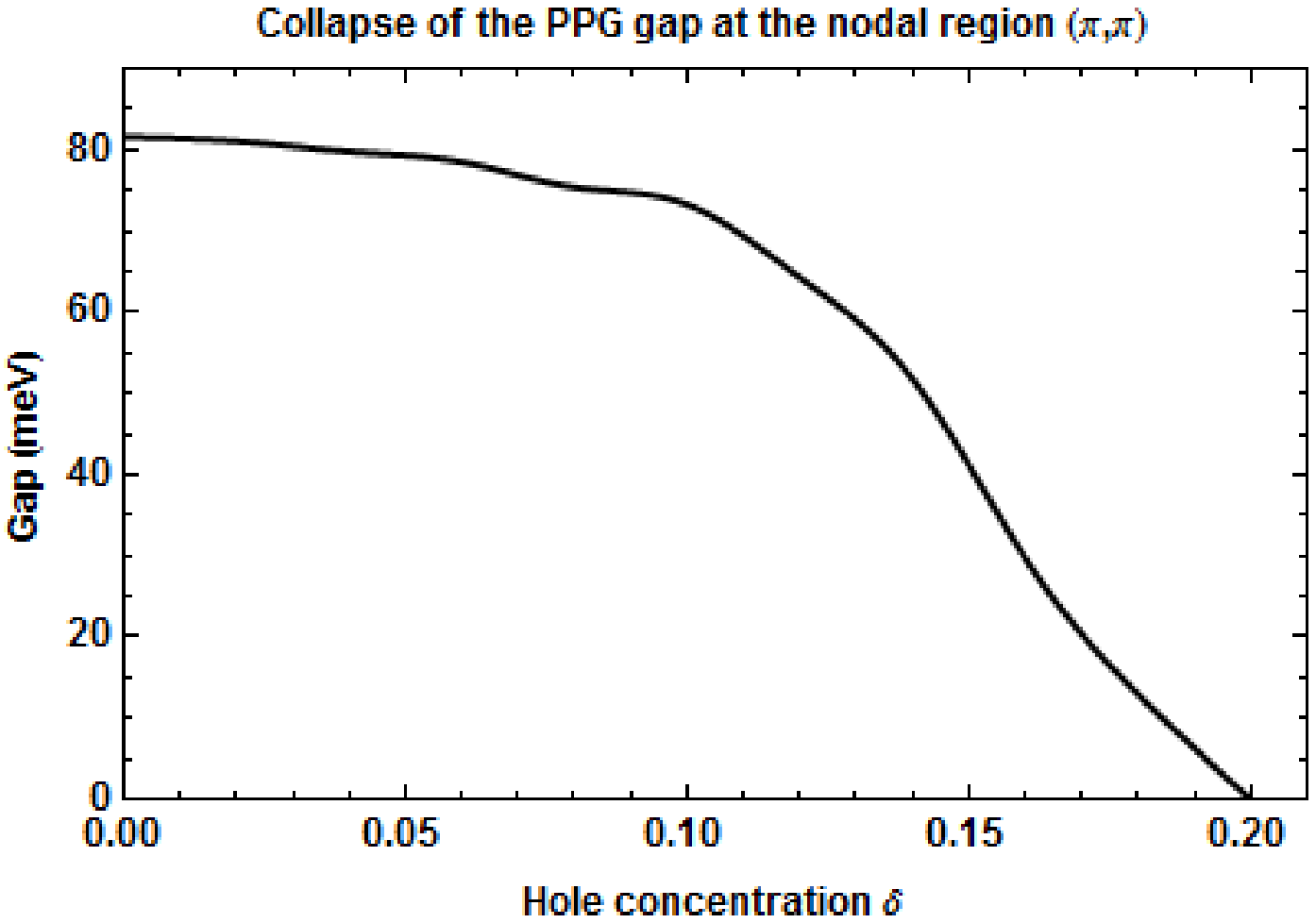}
\par\end{centering}
\caption{ Diminishing of the gap with the increase of the hole concentration. a) For the antiferromagnetic-insulator spectrum. b) For the paramagnetic-pseudogap state. Note that in both cases the gaps are completely closed when the concentration of holes is  $x_{c}\sim0.2$, predicting a metallic behavior.}%
\label{gapafaab}
\end{figure}

The results state that at the moment in which the $AFI$ and  $PPG$ become
degenerated at the quantum critical point,  the obtained $HF$ solution of the
problem becomes a unique metallic and paramagnetic phase. An important point to underline,  is that, as it is shown in figure \ref{gapafaab}, around
the critical doping  $x_{c}=0.2$, the results predict that the insulator gap of
the $AFI$ state diminishes until it  completely collapses. The same occurs to the
state $PPG$. This produces the metallic behavior which coincides with the one
observed in the material in the high doping region.

\section{Evolution of the Fermi surface }

Let us expose now the results for the modifications of the Fermi
surface induced by the hole doping. The Fermi surface, as separating
in the reciprocal space,  the occupied from the empty orbitals
\cite{Kittel}, allows to define many physical properties of the
materials. It is known that  in the largely non understood  normal
state of the $HTc$ superconductors, the Fermi surface becomes
truncated in parts called Fermi arcs. One of the most important
open questions in superconductivity theory is how the Fermi arcs and
the superconductivity are mutually related \cite{Ideta1,Ideta2}.
\begin{figure}[ht]
\begin{centering}
\includegraphics[scale=0.3]{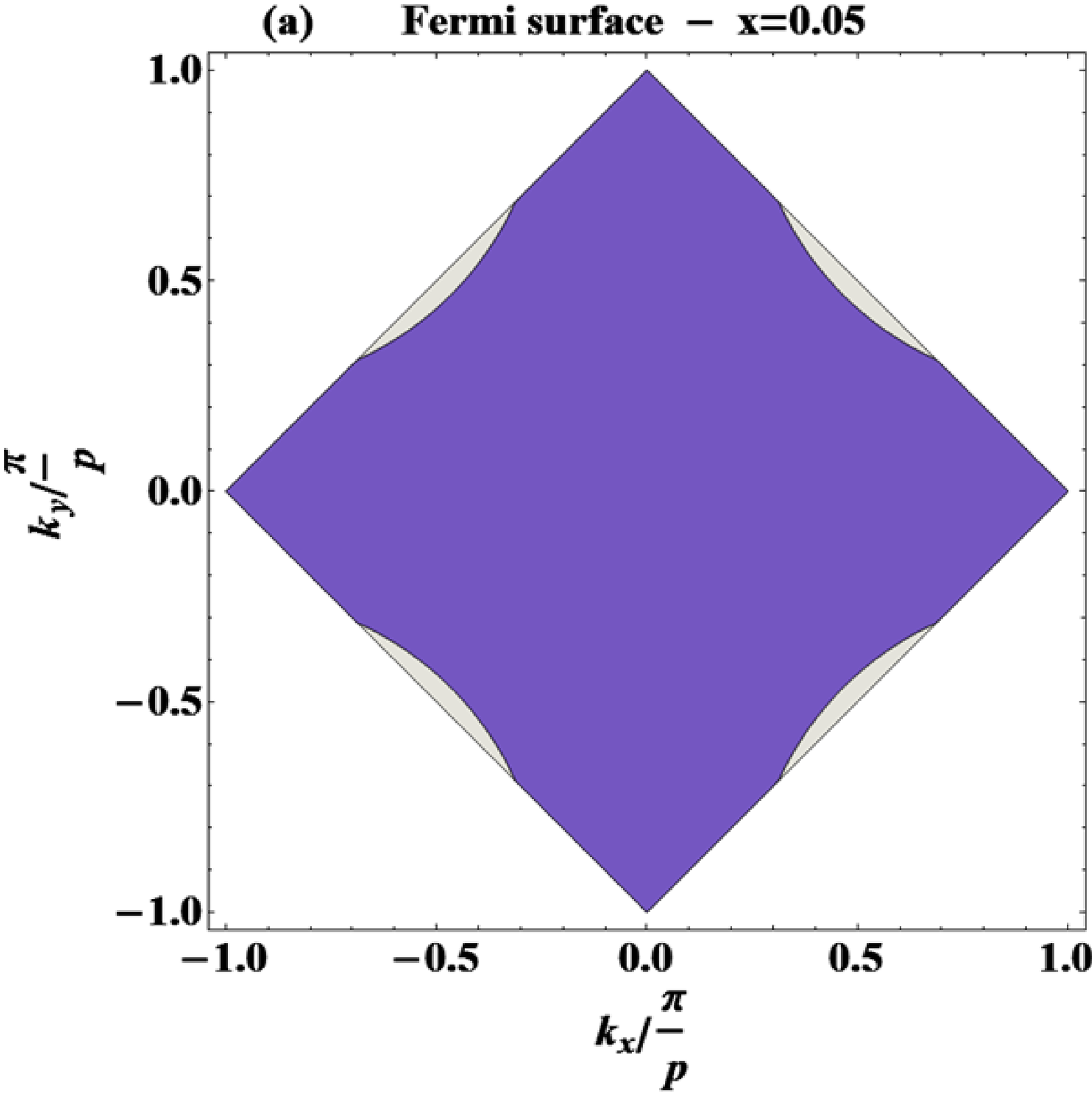}\hspace{8mm}
\includegraphics[scale=0.3]{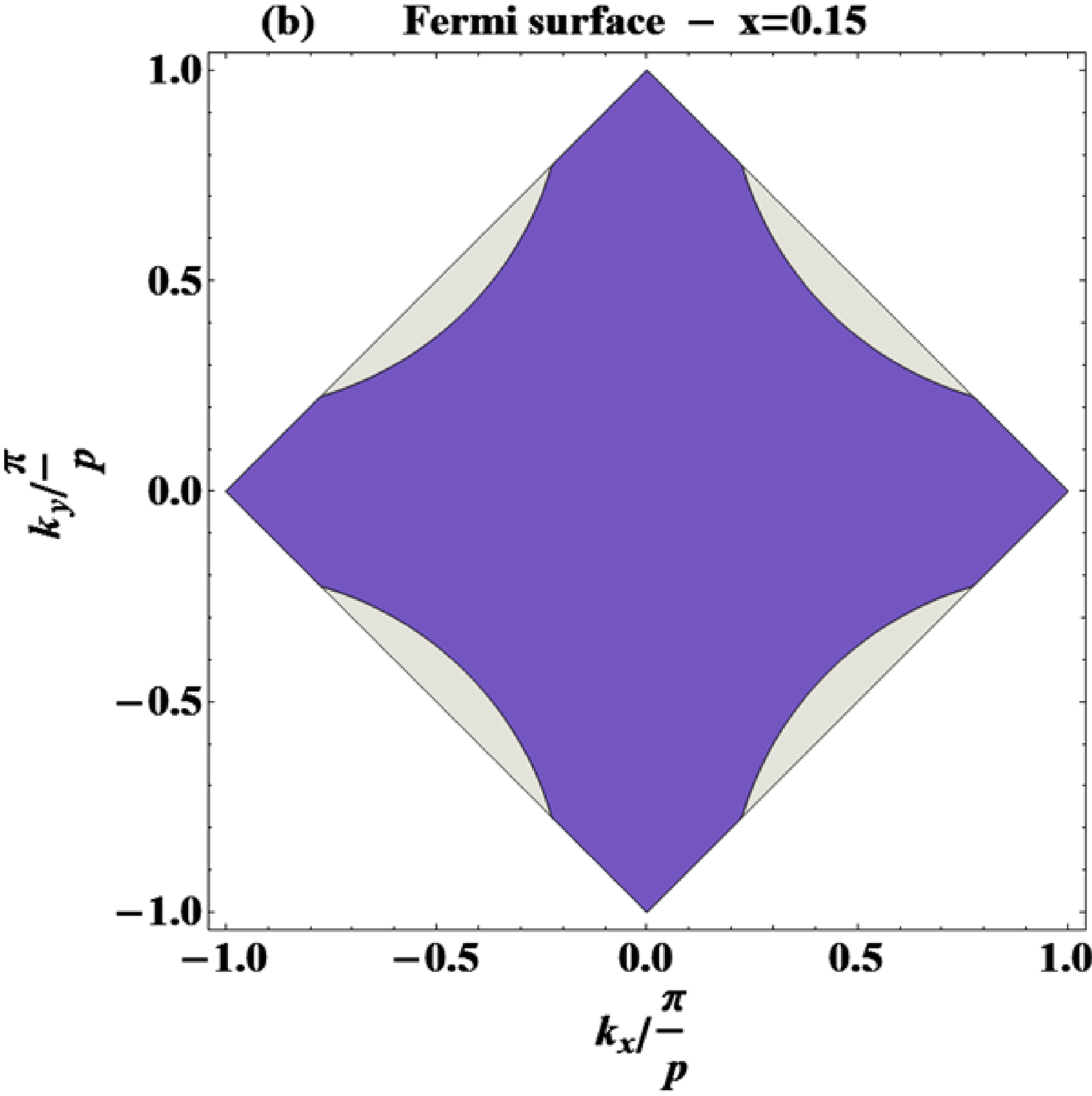}
\par\end{centering}
\par
\begin{centering}
\vspace{3mm}
\par\end{centering}
\par
\begin{centering}
\includegraphics[scale=0.3]{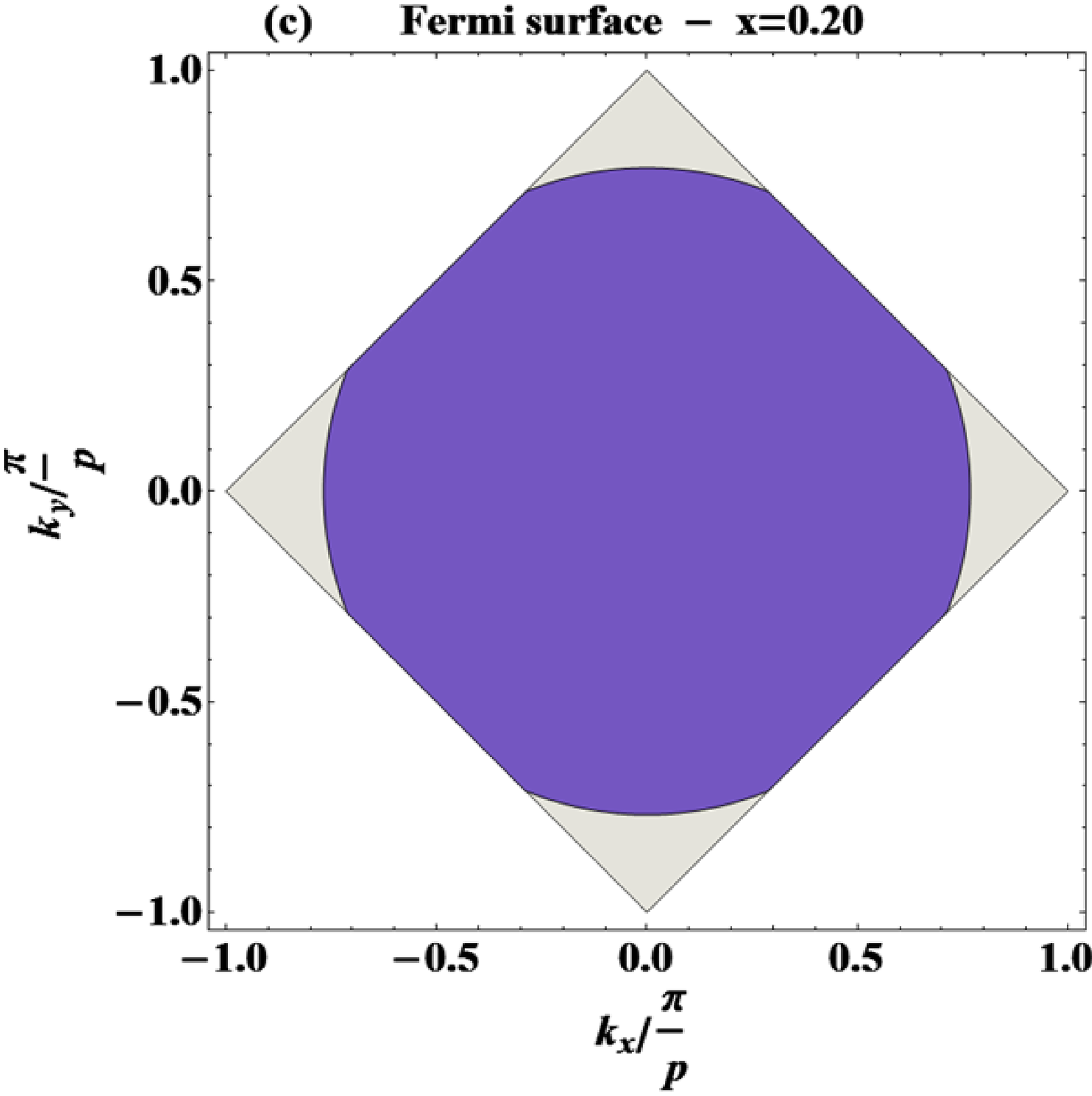}\hspace{8mm}\includegraphics[scale=0.3]{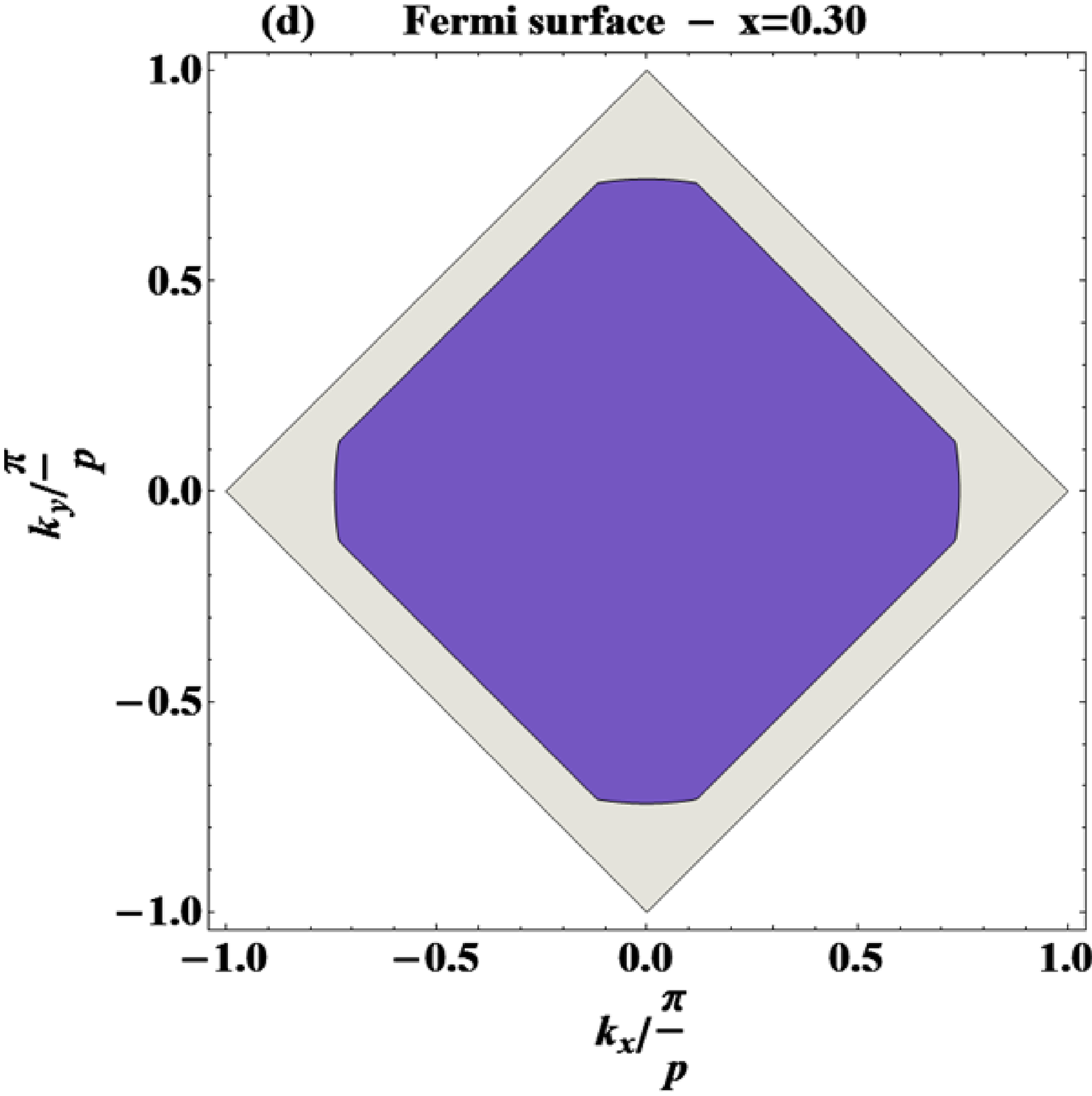}
\par\end{centering}
\caption{ Evolution of the Fermi surface for increasing hole doping.  The areas closed by the Fermi
surfaces are $70$, $80$, $85$ and $95\%$  of the half  Brillouin zone area  and corresponds to $x=0.3$, $0.2$, $0.15$ and $0.05$,
respectively. These proportions are consistent with the Luttinger sum rule, if the electron density is  $1-x$ $=$ ($70$, $80$, $85$ and
$95\%$, respectively).}%
\label{fermisurf}
\end{figure}
We have investigated the Fermi surface predicted by the model and
its dependence on the hole concentration for a wide range of
concentrations $0\leq x\leq0.3$, defining the evolution with doping
from the $AF$-insulator region up to the paramagnetic-metallic  one.

The figure \ref{fermisurf} shows the evolution of the Fermi surface as the hole
doping grows. It can be observed that the model predicts that in the
low doping region ($x\leq0.15$) the  Fermi surface is composed of
the above mentioned Fermi arcs around the  nodal region ($\pi,\pi$). In this direction is  where the doped holes establish as it was mentioned before (the grey zone in figure \ref{fermisurf}). It was
possible to evidence that the length of these arcs grows in
proportion with the doping, up to the attainment of the critical
doping at $x_{c}=0.2,$ where a sort of hole pockets form in
the corners of the reduced Brillouin zone.

In accordance with our calculations shown in figure \ref{fermisurf}, the Fermi surface for $x=0.3$
becomes squared and has a large straight portion around the point
($\frac{\pi}{2},\frac{\pi}{2}$). The areas closed by the Fermi
surfaces are $70$, $80$, $85$ and $95\%$  of the half  Brillouin zone area  and corresponds to $x=0.3$, $0.2$, $0.15$ and $0.05$, respectively. These proportions are consistent with the Luttinger sum rule, if the electron density is  $1-x$ $=$ ($70$, $80$, $85$ and $95\%$, respectively). It can be concluded that our results have a
close qualitative agreement with the ones shown in figure 9, that
were obtained by means of the ARPES in Ref. \cite{Ino}.
\begin{figure}[ht]
\par\begin{centering}
\includegraphics[scale=0.45]{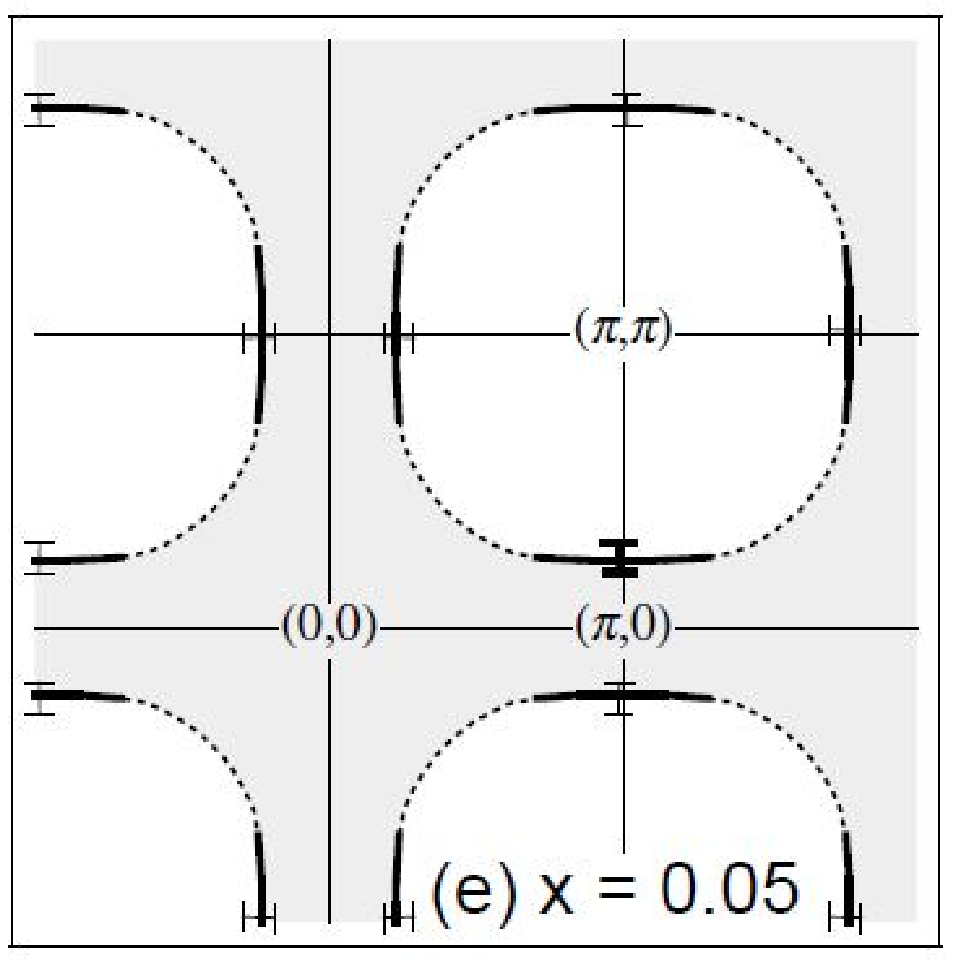}\hspace{8mm}\includegraphics[scale=0.45]{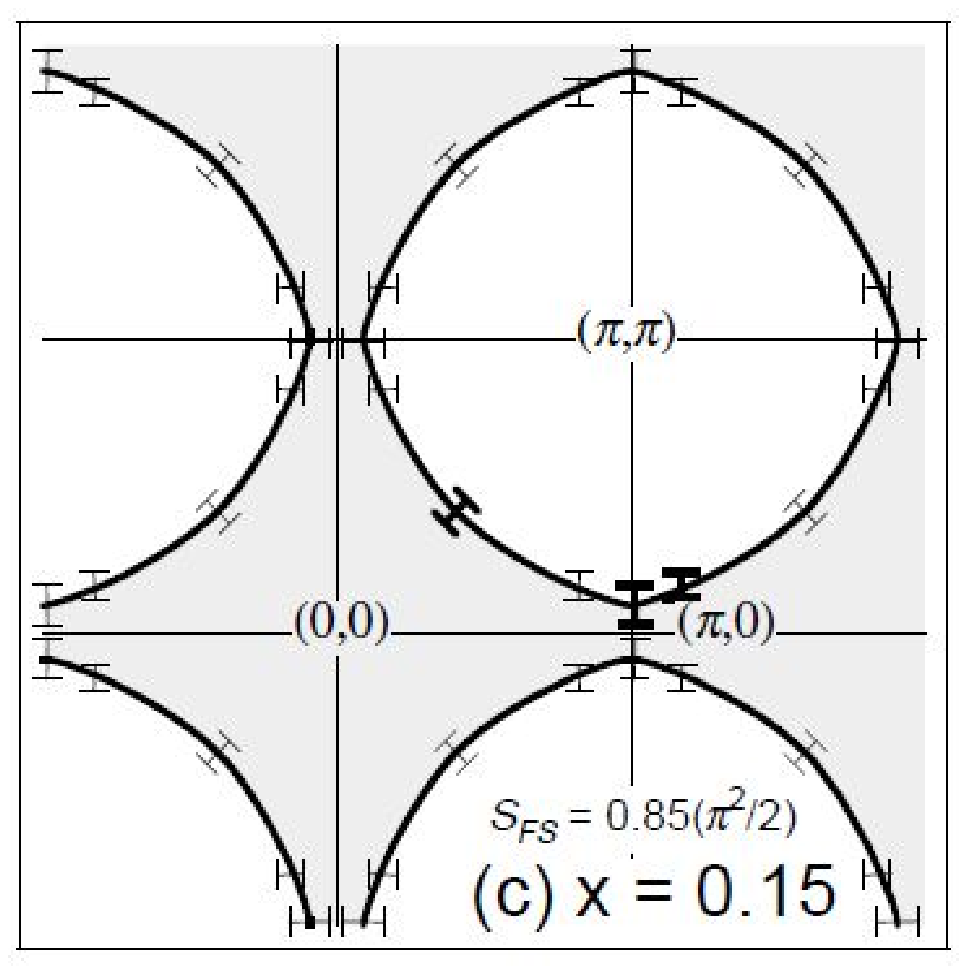}\hspace{8mm}
\includegraphics[scale=0.45]{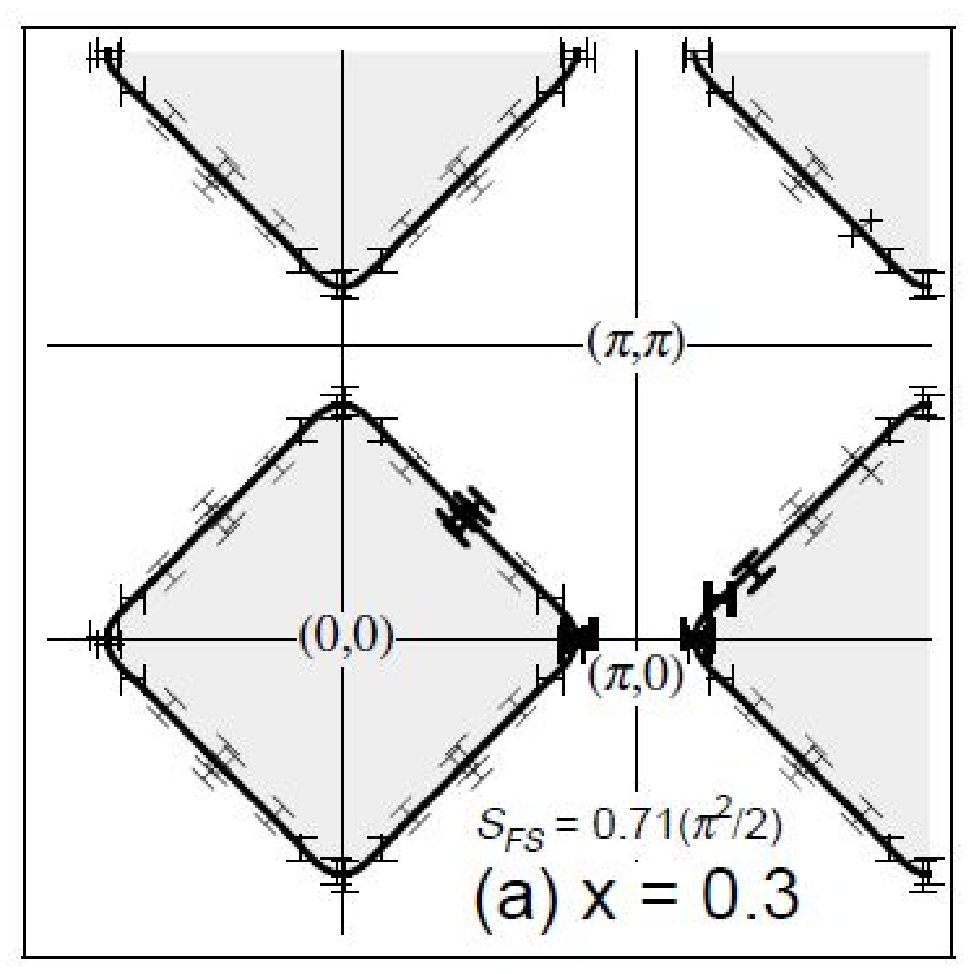}
\par\end{centering}
\caption{   Evolution of the Fermi surface for increasing hole concentration for the  $La_{2-x}Sr_{x}CuO_{4}$,
 obtained by the ARPES experiment.  Taken from: Ino A 1999. The dotted line in the first figure represent the so-called Fermi arc. That is a discontinuous Fermi surface.}%
 \label{fermisurfexp}
\end{figure}

Therefore, it follows  that the considered model predicts that the Fermi surface of the  $LSCO$
undergoes a  transition, which goes
from a hole-like Fermi surface centered at ($\pi,\pi$) for $0<x<0.2$
into an electron-like one centered at ($0,0$) for $0.2<x\leq0.3$. It is believed
that the drastic change of the Fermi surface can correspond with the
fact that the sign of the Hall coefficient changes from positive to
negative values around  $x=0.25$ in the  $LSCO$ \cite{Tamasaku,Uchida}. In the
framework of the investigated model,  the mentioned change is
associated to the occurrence of a quantum phase transition around
the critical point  $x_{c}=0.2$ which lay beneath the
superconductor dome.

\section{The quantum phase transition }

One of the  current questions in $HTc$ superconductor theory
is the occurrence of the superconductivity as determined by the
existence of a quantum critical point laying  within the hole
doping interval at which the superconductivity occurs. Today is widely
debated the question about what is the detailed connection between
the critical point and the superconductivity effect. Experimental indications
about the existence of a  critical quantum point comes from the transport and thermodynamic measurements \cite{Balakirev,Daou,Marel,Bernhard}. In the
framework of the present study,  the $HF$
energy per particle for the  $AFI$ and $PPG$ states were evaluated
by varying the hole concentration in the range $0\leq x\leq0.3$, by
using the formula
\[
E_{\mathbf{k},l}^{HF}=\sum_{\mathbf{k},l}\Theta_{(\tilde{\varepsilon}%
_{F}-\tilde{\varepsilon}_{l}(\mathbf{k}))}[\tilde{\varepsilon}_{l}%
(\mathbf{k})-\frac{\tilde{\chi}}{2}B^{\mathbf{k},l\ast}.(G_{\mathbf{k}}%
^{C}-G_{\mathbf{k}}^{i}).B^{\mathbf{k},l}].
\]

The figure 10 (a) show how the $AFI$ state, which have the lowest
energy at half-filling,   evolves and becomes degenerate with the
$PPG$ state at the critical doping $x_{c}=0.2$. At this point, as it
can be seen in the figure 6,  the results predict for the
$AFI$ state, that the insulator  gap diminishes  to be completely
closed. The same behavior is shown by  the  $PPG$ state for which the momentum
dependent pseudogap collapses.  The solution of the problem determines a metallic behavior
which is the observed nature of $La{_2}CuO{_4}$  for high hole
concentrations.

\begin{figure}[ht]
\par\begin{centering}
{\scriptsize
(a)}\includegraphics[scale=0.5]{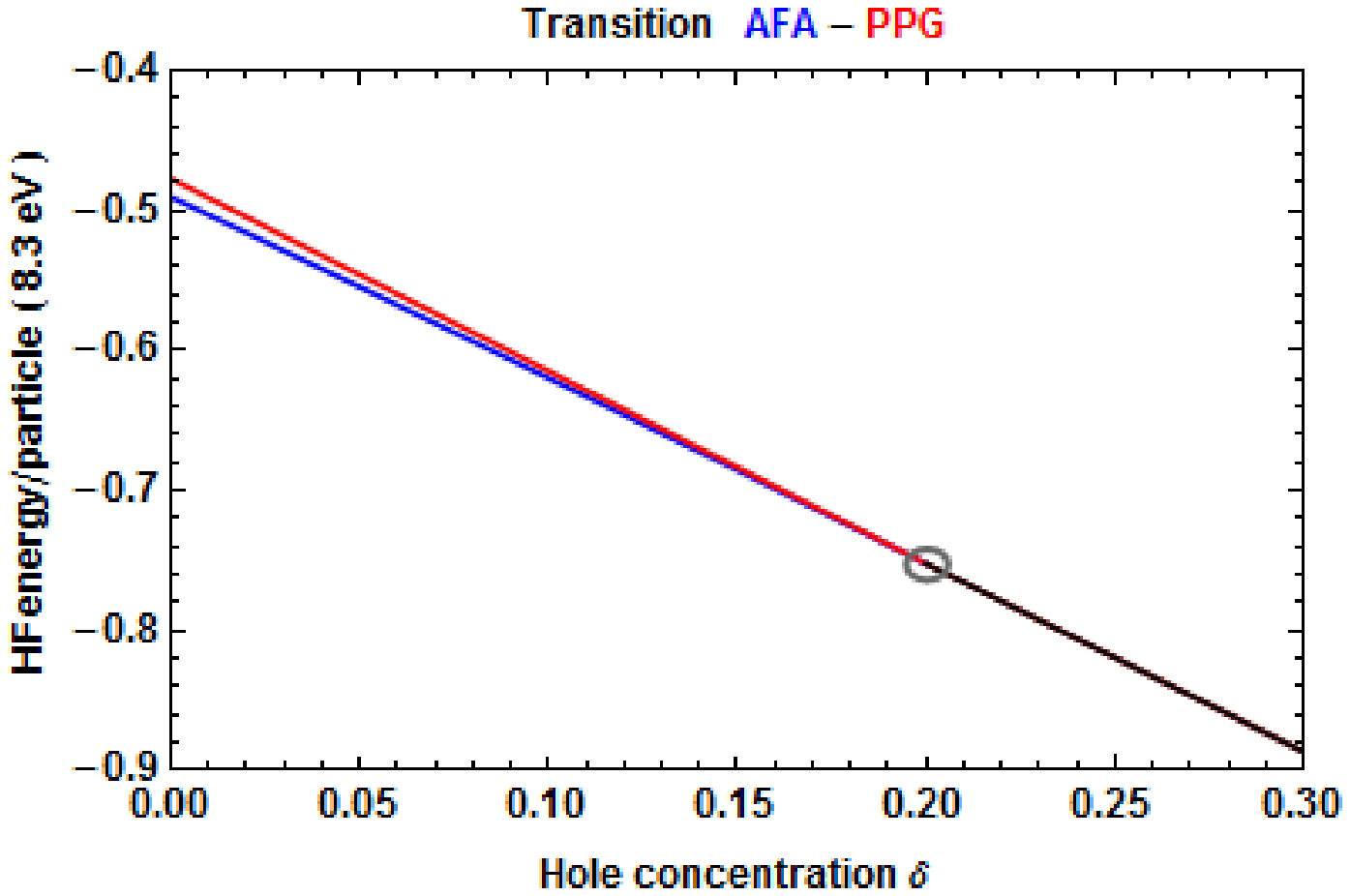}\hspace{8mm}
\hspace{8mm}{\scriptsize
(b)}\includegraphics[scale=0.3]{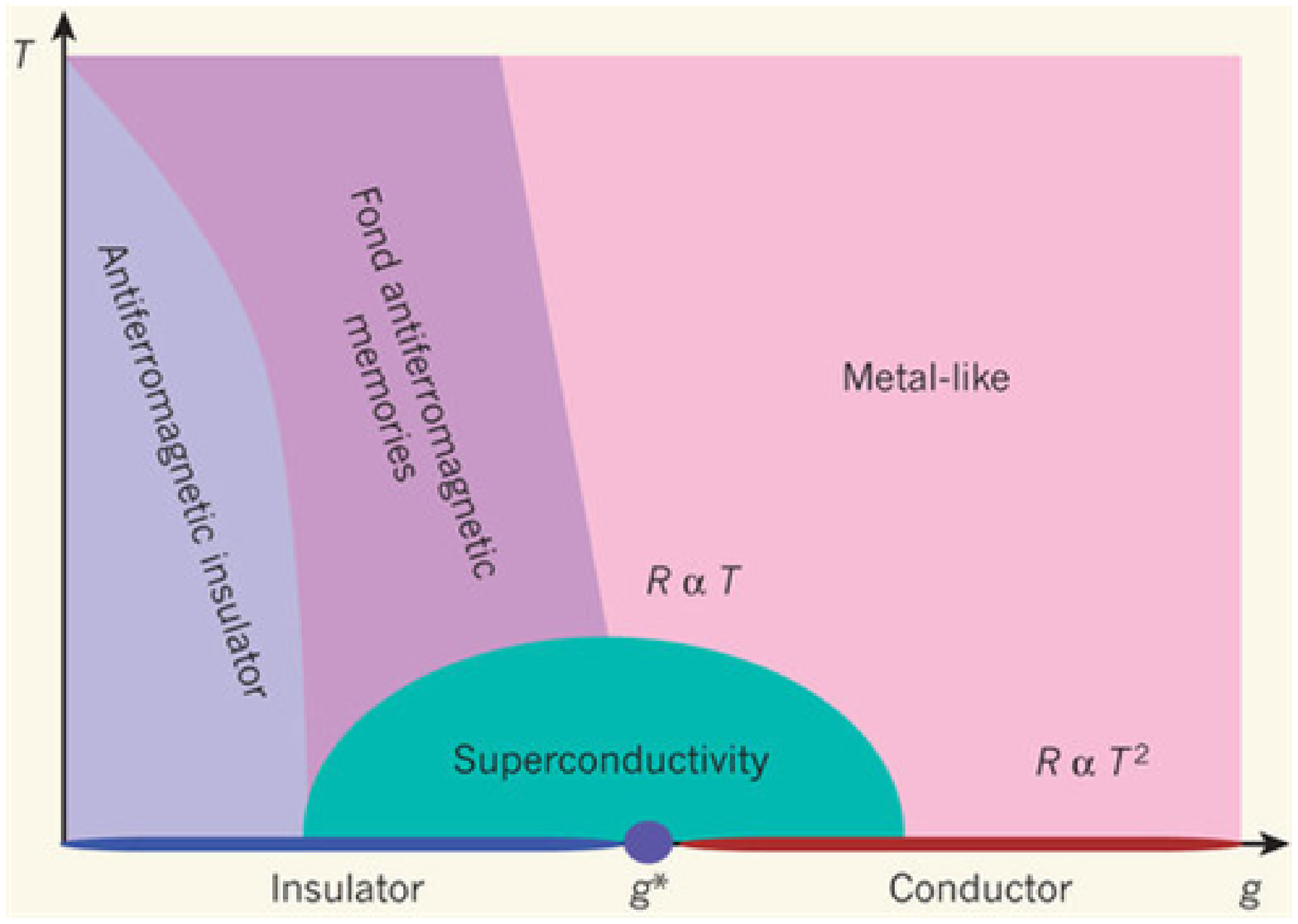}
\par\end{centering}
\caption{{\protect (a) Dependence of the $HF$ energy per particle with the hole doping for the $AFI$ and $PPG$ states. (b) Generic phase diagram where it can be observed the quantum phase transition from the $AFI$ state to a paramagnetic-conductor.  }}%
\label{transition}
\end{figure}
\begin{figure}[ht]
\begin{centering}
\includegraphics[scale=0.40]{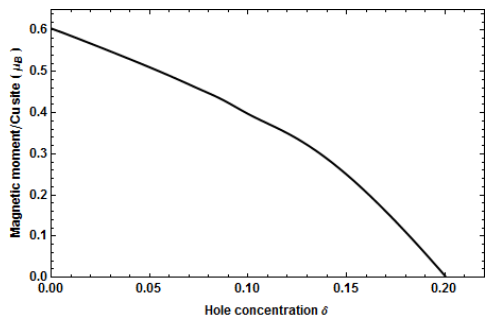}
\par\end{centering}
\caption{{\protect Reduction of the magnetic moment of the $AFI$ state under the doping increasing.}}%
\label{magneticmom}
\end{figure}

The magnetization of the $AFI$ state was also evaluated by its defining formula

\begin{equation}
\mathbf{m}(\mathbf{x})=\sum_{\mathbf{k},l}\sum_{s,s^{\prime}}\phi
_{\mathbf{k},l}^{\ast}(\mathbf{x},s)\boldsymbol{\sigma}(s,s^{\prime}
)\phi_{\mathbf{k},l}(\mathbf{x},s^{\prime}),
\label{eq:25}
\end{equation}
where

\begin{equation}
\boldsymbol{\sigma}(s,s^{\prime})=\sigma_{x_{1}}(s,s^{\prime}%
)\mathbf{e}_{x_{1}}+\sigma_{x_{2}}(s,s^{\prime})\mathbf{e}{}_{x_{2}}%
+\sigma_{z}(s,s^{\prime})\mathbf{e}_{z},\label{eq:26}%
\end{equation}
and $\sigma_{x_{1}}$, $\sigma_{x_{2}}$ and $\sigma_{z}$ are the Pauli matrices. One
important conclusion in this study is that in approaching  the
critical doping value the local magnetization of the $AFI$ state
tends to vanish (see figure \ref{magneticmom}), a property that the $PPG$ showed
along its whole hole doping evolution from the initial half-filled
state.

\section{A new hole binding mechanism: spin-space
entanglement }

The evaluation procedure  of the energy for the $AFI$ state  in the
region of low doping showed a clear dependence on the odd or even
character of the number of holes added. That is, it was needed a
very much larger number of iterations for attaining  convergence
when the number of holes was odd.

We estimate that a reason for such a behavior might be that adding
an odd number of holes, could be expected to show a larger complexity of the
$HF$ single particle states, since the Kramers degeneracy should be
implemented. That is, it could be expected that the double
degeneracy implied by the time inversion invariance of the system,
when the odd number of holes does not allow to fill an integer
number of degenerate pairs of states, should make such single
particle states to show a very much complex structure.

 A perhaps related circumstance is that the  spin-spatial entangled
 structure of the $HF$ single particle states leads to imagine a  mechanism of hole pairing that
could be acting in the considered system. The idea is that this
entangled structure of the single particle states might
produce that two, well separated in space,  hole wave packets could show,
each of them, their  relatively complex spin-spatial entangled
constitution. However, when the two holes are allowed to be close in
space, it is possible  that they could tend to compensate their
spatially dependent spin and magnetic moment structures and show a
lower energy than the other pair of  well separated holes. In the case of the
paramagnetic systems such an effect seem  to be very much
weaker, due to the simplicity of the spin structure.

In order to check for this possibility we defined a quantitative
measure of the binding between two holes in the antiferromagnetic
state. This  definition was based in the energy of the ground
state of  the system $E_{0}$ at half-filling and the ground state
energies after doped with one hole $E_{1}$ and with two holes
$E_{2}$ in the form
\[
\triangle_{B}=e_{2}-2e_{1},
\]
where $e_{1}=E_{1}-E_{0}$ and $e_{2}=E_{2}-E_{0}$. Whenever,  two
holes minimize their energy by producing a bound state, then
$\triangle_{B}$ becomes negative. When $\triangle_{B}$ vanishes the
holes may not form a bound state since then $e_{2}=2e_{1}$ and it is
expected that they behave as independent excitations.

\begin{figure}[ht]
\par\begin{centering}
\includegraphics[scale=0.42]{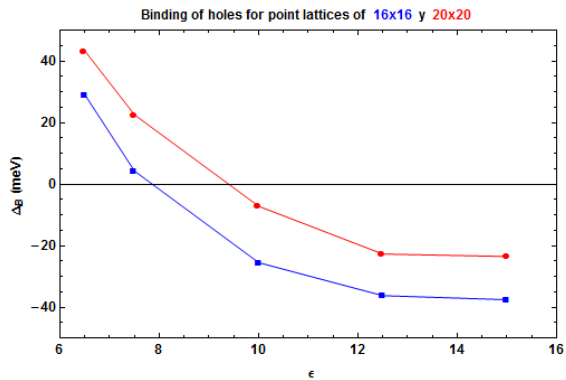}
\par\end{centering}
\caption{ {\protect Binding energy  $\triangle_{B}$ of two holes as
a function of the dielectric constant  $\epsilon$ for point lattices
of $16\times16$ (blue squares) and $20\times20$ (red circles).}}%
\label{binding}
\end{figure}

The binding energy of two holes $\triangle_{B}$ as a function of the
dielectric constant $\epsilon$ for point lattices of $16\times16$
sites (squares) and $20\times20$ sites (circles) in the
case of the $AFI$ state is shown in figure \ref{binding}. Note that the bound
states of the two holes form at low doping value as helped by the
amount of dielectric screening of the Coulomb interaction. The
figure \ref{binding} indicates that in general, when the dielectric constant
$\epsilon$ increases, the binding energy reduces until they form a
bound state. The critical value for the appearance of binding for
the holes slowly grow with the increasing of the size of the sample
(the region in which periodic boundary conditions were imposed).

The results indicate that although the thermodynamical limit is not yet
attained for the evaluation of this quantity $\triangle_{B},$ it seemingly
exists. An interesting result is that
in $La_{2}CuO_{4}$,  it has experimentally determined that a
superconductor gap $\triangle_{SC}$ is the order of the $10\,meV$ \cite{Ino},
which is of the similar magnitude  of the shown values in figure
\ref{binding},  for the two holes  binding energy. That  is, the measured
superconductor gap approximately coincides with the energies
$\triangle_{B}$ required to break  the two bound holes. This outcome
suggests the possibility that the $HF$ solution  is able to also
convey a pairing effect determining the presence of $HTc$
superconductivity  and the constitution of the  Cooper
pairs.

A possibly acting  binding mechanism  could be as  follows. Firstly,
recall that the  $HF$ single particle states show the more
complicate than the standard  spin-spatial entangled structure.
This complex composition can be expect to provoke, if the screening is
assumed to be strongly enough, a diminishing energy effect, in which
the  respective entangled magnetic moment structure of the
combined two  states  could tend to compensate one to another. Such
an effect should appear in the  bound state Bethe-Salpeter
equations for two holes. In the described picture, the bound state
of two holes at low doping in the $AFI$ phase will be formed in this
picture thanks to the screening of the Coulomb interaction. The fact that the
measured static dielectric constant of the $La_2CuO_4$ is around the value 25,
 indicates that the amount of screening is high.

In the more investigated models, like the  $t-J$ one, it is argued
that the existence of each hole breaks four antiferromagnetic bounds
\cite{Dagotto}, an effect that have an energy cost of the order of the
coupling energy  ($0.1\,eV$). Therefore, at least in the  low hole
density limit, like the case under consideration, two holes minimize
the energy to create them by sharing a common bound. In this way
they minimize the number of broken antiferromagnetic bounds. Note that this picture
seems to be  compatible with the one previously exposed on the basis of the spin-spatial
entangled nature of the hole states. It is also known that the size of  the
Cooper pairs of the material is  a small quantity (around two lattice
constants) \cite{Dagotto}.

It had been also argued that the existence of  preformed hole
pairs could not be sufficient evidence for the appearance of the superconductivity. In the context of the present model, it has been argued  that   for large values the dielectric constant
and the presence of spin-spatial entanglement,  the pairs could be formed.  Then, let us assume that their pair wave function  can be formed by  superposing  products of single particle waves having  momenta values  close to the  centers of the four  reduced Brillouin zone sides in figure 6.  Then,  the amount of the momentum transfer associated to these functions in the Bethe-Salpeter equation could be expected to produce  pair wave functions showing  sizes  of the order of  few lattice cells.  If such results to be the situation, then,  the hole pairs could be expected to condense at $T=0\,K$ giving   rise to a Bose condensate of Cooper pairs showing
superconductivity. These possibilities will be investigated elsewhere. In ending this section,  we also want to express that the results also suggests that the here argued binding effects between two added holes, could  constitute
a dynamical foundation  of the physical meaning of the charge two boson excitations  identified in Ref. \cite{Philips},  as being relevant for the description of Hubbard models.

\section{Conclusions}

We  have investigated  the model for the $La_{2}CuO_{4}$
defined  in Refs. \cite{Cabo1,Cabo2,Cabo3} applied  to the
situation in which the material is doped with holes. The main
elements of the model were reviewed by fixing its free parameters
and solving the Hartree-Fock equations. The evolution with doping of
the band spectrum of the antiferromagnetic-insulator state and the
metallic-paramagnetic pseudogap one are presented for a wide range of
hole concentrations $0\leq x\leq0.3$.  Around the critical doping
density  $x_{c}=0.2$, the results show that for the $AFI$ state,
the band spectrum suffers a change in structure of the bands and the
insulating band  gap diminishes  until its complete closing.  A
similar  process also occurs for the $PPG$ state where the
pseudogap collapses at the same special hole density. Over this critical density the system only
shows a single paramagnetic-metallic state. The magnetization of the
$AFI$ state becomes zero exactly at the critical point. The
solutions evidence that the destruction of the antiferromagnetic
order is produced by the fact that added holes tend to occupy the
states which show a more intense $AF$ order: that ones which are
closer to the boundary of the reduced Brillouin zone. The results
show a drastic change of the nature of the Fermi surface, which goes
from a hole-like Fermi surface centered at ($\pi,\pi$) for $0<x<0.2$
into an electron-like one centered at ($0,0$) for $0.2<x\leq0.3$.
Henceforth, the whole discussion  had shown that the investigated
model predicts the existence of a quantum phase transition at
critical doping value which is beneath the superconducting dome in
$La_{2}CuO_{4}$. It was also identified a possibility for the
pairing of holes which could give rise to the superconductivity and the bounding energy was estimated. The
effect seems to be closely linked to the spin-space
\textquotedblleft entanglement\textquotedblright effect of the $HF$
single particle states. Possible connections with the existence of bosonic doubly charged excitations investigated in Ref. \cite{Philips} were also pointed out.

In ending, let us comment on some possible extensions of the present
work. It seems convenient to consider a new parameter for to be fixed in the
Hamiltonian of the model. It is related with the effective mass
defining the kinetic term of the free part of  Hamiltonian. This
constant was assumed to be equal to the electron mass in free space,
which,  possibly,  is a somewhat rude choice, since such electrons are
assumed to move in the crystalline potential generated by all the
other  particles, filling the rest of the many bands. This new freedom
in the parameters might be relevant when the spin-orbit effects will be
taken into account in order to describe the magnetic anisotropy of the
$AF$ order in further discussions. It  would be also of
help in allowing to more precisely fix the value of the insulator
gap of the $AFI$ state which is known to be close to $2\,eV$. One important point in this respect
is that the dielectric constant $\epsilon$ has been measured for $La_{2}CuO_{4}.$
Therefore, it will be needed to fix the free  parameters in order to  define the observed
value of the gap. At this point it can be noticed that electron effective mass will be an influential
value which can control  the energy scale of the bands and could  be
phenomenologically fixed to define the $2\,eV$ insulator gap. It can be observed from equations (\ref{eq:17}) and (\ref{eq:18}).

It seems also of interest to perform calculations in order to attain
the thermodynamic limit in the hole pairing effect, and also to
attempt deriving this binding effect but in the framework of the
Bethe-Salpeter equation for two holes moving in the medium. This would
confirm the presence of preformed Cooper pairs in the $AFI$ state of
the model. Lastly, it will seem convenient to improve the formulation of the
model by employing 3D Wannier states resembling the incomplete 3D
shell of the $Cu$ atoms within the $CuO_{2}$ planes. Upon this, the
incorporation of the spin-orbit interactions will be needed in order
to further describe the magnetic anisotropy of the $AF$ order in
$La_{2}CuO_{4}$.

\begin{acknowledgments}
The authors would like to acknowledge the helpful remarks of Profs. E. C. Marino, A. LeClair, A. Gonz\'alez,
E. Altshuler, A. Odriazola and J. C. Sua\'arez. We are also indebted by the helpful support received from
various institutions: the Caribbean Network on Quantum Mechanics, Particles
and Fields (Net-35) of the ICTP Office of External Activities (OEA).

\end{acknowledgments}

\end{document}